\documentclass[conference]{IEEEtran}

\ifCLASSINFOpdf
\else
\fi

\usepackage{siiscmds}                           %
\usepackage{siunitx}
\usepackage{setspace}
\usepackage{caption}
\usepackage{subcaption}
\usepackage{lipsum}
\usepackage{pifont}
\usepackage{booktabs}
\usepackage{amsfonts}
\usepackage{hyperref}
\usepackage[style=ieee]{biblatex}
\usepackage{pbalance}
\usepackage{tabularx}
\usepackage{mdframed}
\newcommand{\toolname}{\textsc{EIPSim}\xspace}
\newcommand{\clusterdata}{\texttt{clusterdata-2019}\xspace}

\newcommand{\Random}{\textsc{Random}\xspace}
\newcommand{\LRU}{\textsc{LRU}\xspace}
\newcommand{\Tagged}{\textsc{Tagged}\xspace}
\newcommand{\Segmented}{\textsc{Segmented}\xspace}

\SetKwInput{KwInput}{Input}
\SetKwInput{KwOutput}{Output} 
\SetKwProg{Fn}{Function}{ :}{end}
\SetKw{Ret}{return}
\SetKwComment{Comment}{/* }{ */}
\SetKwFunction{REL}{RELEASE}
\SetKwFunction{INIT}{INIT}
\SetKwFunction{PRALL}{RANDOM.ALLOCATE}
\SetKwFunction{PRREL}{RANDOM.RELEASE}
\SetKwFunction{PRINIT}{RANDOM.INIT}
\SetKwFunction{LRUALL}{LRU.ALLOCATE}
\SetKwFunction{LRUREL}{LRU.RELEASE}
\SetKwFunction{LRUINIT}{LRU.INIT}
\SetKwFunction{IPTAGGINGALL}{TAGGED.ALLOCATE}
\SetKwFunction{IPTAGGINGREL}{TAGGED.RELEASE}
\SetKwFunction{IPTAGGINGINIT}{TAGGED.INIT}
\SetKwFunction{IPSCANSEGMENTATIONALL}{SEGMENTED.ALLOCATE}
\SetKwFunction{IPSCANSEGMENTATIONREL}{SEGMENTED.RELEASE}
\SetKwFunction{IPSCANSEGMENTATIONINIT}{SEGMENTED.INIT}
\SetKwFunction{setIpAllocated}{setIpAllocated}
\SetKwFunction{setIpNotAllocated}{setIpNotAllocated}
\SetKwFunction{randomSample}{randomSample}
\SetKwFunction{createIp}{createIp}
\SetKwFunction{getCurrentTime}{currentTime}
\SetKwFunction{getIpClosestDurationTracker}{getIpClosestDurationTracker}
\SetKwFunction{findTenant}{findTenant}

\renewcommand{\paragraph}{\vspace{4pt}\textbf}

\usepackage[available,functional,reproduced]{ndssbadges}

\newcommand{\bestsynmtadvsegmentednewLatentConfsImprovementOvertagged}{\SI{70.1}{\%}}
\newcommand{\bestsynmtadvsegmentednewLatentConfsImprovementOverrandom}{\SI{83.8}{\%}}
\newcommand{\bestsynmtadvtaggednewLatentConfsImprovementOverrandom}{\SI{46.0}{\%}}

\newcommand{\segmentedLCYieldVariation}{$40\%$}
\newcommand{\segmentedIPYieldVariation}{$25\%$}

\begin{document}
\title{Secure IP Address Allocation at Cloud Scale}

\author{\IEEEauthorblockN{Eric Pauley\IEEEauthorrefmark{1}\IEEEauthorrefmark{3},
Kyle Domico\IEEEauthorrefmark{1},
Blaine Hoak\IEEEauthorrefmark{1},
Ryan Sheatsley\IEEEauthorrefmark{1},
Quinn Burke\IEEEauthorrefmark{1},\\
Yohan Beugin\IEEEauthorrefmark{1},
Engin Kirda\IEEEauthorrefmark{2},
Patrick McDaniel\IEEEauthorrefmark{1}}
\IEEEauthorblockA{\IEEEauthorrefmark{1}University of Wisconsin--Madison}
\IEEEauthorblockA{\IEEEauthorrefmark{3}Email: epauley@cs.wisc.edu}
\IEEEauthorblockA{\IEEEauthorrefmark{2}Northeastern University}}

\IEEEoverridecommandlockouts
\makeatletter\def\@IEEEpubidpullup{3.5\baselineskip}\makeatother
\IEEEpubid{\parbox{\columnwidth}{
    Network and Distributed System Security (NDSS) Symposium 2025 \\
    23 - 28 February 2025, San Diego, CA, USA \\
    ISBN 979-8-9894372-8-3 \\
    https://dx.doi.org/10.14722/ndss.2025.23374\\
    www.ndss-symposium.org
}
\hspace{\columnsep}\makebox[\columnwidth]{}}

\maketitle

\begin{abstract}
Public clouds necessitate dynamic resource allocation and sharing. However, the dynamic allocation of IP addresses can be abused by adversaries to source malicious traffic, bypass rate limiting systems, and even capture traffic intended for other cloud tenants. As a result, both the cloud provider and their customers are put at risk, and defending against these threats requires a rigorous analysis of tenant behavior, adversarial strategies, and cloud provider policies. In this paper, we develop a practical defense for IP address allocation through such an analysis. We first develop a statistical model of cloud tenant deployment behavior based on literature and measurement of deployed systems. Through this, we analyze IP allocation policies under existing and novel threat models. In response to our stronger proposed threat model, we design \textit{IP scan segmentation}, an IP allocation policy that protects the address pool against adversarial scanning even when an adversary is not limited by number of cloud tenants. Through empirical evaluation on both synthetic and real-world allocation traces, we show that IP scan segmentation reduces adversaries' ability to rapidly allocate addresses, protecting both address space reputation and cloud tenant data. In this way, we show that principled analysis and implementation of cloud IP address allocation can lead to substantial security gains for tenants and their users.

\end{abstract}

\section{Introduction}\label{sec:introduction}
Cloud providers allow near limitless scalability to tenants while reducing or eliminating upfront costs. One component that enables this architecture is the reuse of scarce IPv4 addresses across tenants as services scale. Though a practical necessity, this reuse--combined with the use of IP addresses as a \textit{security principal}--enables malicious cloud tenants to abuse IP address reputation~\cite{ioannidis_implementing_2000, sinha_shades_2008, zhang_highly_2008}, pollute the address space for future tenants~\cite{antonakakis_building_2010}, and even collect sensitive information intended for previous tenants~\cite{liu_all_2016, borgolte_cloud_2018, pauley_measuring_2022}. We observe that these seemingly disparate attack spaces share a common thread: the ability of adversaries to easily sample large numbers of IP addresses from provider pools.

While prior works have identified and confirmed the issue of IP address reuse, and proposed some preliminary mitigations~\cite{borgolte_cloud_2018, pauley_measuring_2022}, the community still lacks a complete understanding of the security provided by these measures, especially against a more powerful or adaptive adversary. For instance, prior works that attempt to reassign addresses to the same tenant can be defeated by adversaries using many disconnected cloud accounts (a form of Sybil attack). Developing secure policies for IP address allocation necessitates a fine-grained analysis of tenant behaviors and adversarial strategies. Such an analysis, and the stronger defenses that analysis enables, are the key focus of this work.

Towards this goal, we propose a novel, comprehensive model for IP address allocation on public clouds. By considering realistic distributions of benign tenant behaviors, configuration management, and cloud provider allocation policies, our new model enables us to concretely evaluate the effectiveness of attacks against the address pool. Implemented in the Elastic IP Simulator (\toolname), tenant and adversarial behaviors enable the key goal of our work: developing new allocation strategies that reduce the ability of adversaries to allocate, measure, and exploit many IP addresses. Our model is validated via real-world data on cloud tenant allocations, as well as data collected on cloud configuration management practices and discussions with major cloud providers. In this way, our model enables the development of new defenses against a broad class of attacks against cloud services.

Our model enables us to characterize and defend against a stronger adversary than considered in prior work. This \textit{adaptive adversary} performs a Sybil attack against the cloud provider, creating many accounts to continually allocate new IP addresses from the pool. Hence, this attacker effectively defeats the protections provided by prior works. We propose \textit{IP scan segmentation}, a novel IP allocation policy that heuristically identifies adversarial behavior across many cloud tenants, and effectively segments the pool to prevent such adversaries from allocating many unique IPs and exploiting vulnerabilities.

We use \toolname to evaluate the security properties (i.e., adversarial ability to discover unique IPs and exploitable configurations) of our studied allocation policies and tenant/adversarial behaviors in real cloud settings. Our analysis, spanning over 250 years of simulated IP address allocation, highlights the marked impact of IP allocation policies on the exploitability of IP address reuse. Indeed, our analysis shows that IP scan segmentation reduces adversarial success by \bestsynmtadvsegmentednewLatentConfsImprovementOverrandom{} over the IP allocation policies deployed by cloud providers, and by \bestsynmtadvsegmentednewLatentConfsImprovementOvertagged{} compared to prior explored techniques. Because our model concretely parallels the actual behavior of cloud providers and tenants, the techniques studied in this work can be directly implemented by providers to protect their customers and network resources. We have shared our findings with providers and release our models and policies as open source artifacts\footnote{\url{https://github.com/MadSP-McDaniel/eipsim/}} to support practical security of IP address allocation.

IP address reuse poses a practical security concern, but principled study of new allocation techniques can lead to practical defenses, making this reuse less exploitable in practice. Our work provides such a defense, as well as a basis on which future research in IP allocation can be measured.

{}

\section{Background}\label{sec:background}

Our work addresses security properties of IP address allocation for public clouds. As such, we briefly describe considerations in IP allocation generally, as well as contemporary work in cloud security related to IP address allocation.

\subsection{IP Address Allocation}\label{sec:background:allocation}

Network hosts require an IP address for communication. This can be manually assigned or managed out of band, or it can be provisioned through some automation. In home and corporate networks, the standard solution to automatic IP allocation is DHCP~\cite{droms_dynamic_1997}. Likewise, in public clouds such as Amazon Web Services~\cite{aws_website}, Microsoft Azure~\cite{azure_website} or Google Cloud~\cite{google_cloud_website}, servers are allocated a private (\ie RFC1918~\cite{moskowitz_address_1996}) IP address via DHCP~\cite{droms_dynamic_1997}. While the DHCP standard does not specify how addresses are assigned, they are generally drawn from a pool either sequentially or based on the physical (MAC) address of the requesting machine~\cite{droms_dynamic_1997}. For workloads with only private or outbound communications, these addresses are sufficient, as outbound connections can be mapped to publicly-routable IPs via Network Address Translation (NAT)~\cite{holdrege_ip_1999}.

When services need to receive connections from the broader Internet, they require a public IP address (usually, at a minimum, an IPv4, though support is increasing for IPv6~\cite{noauthor_ipv6_nodate}). These addresses could be configured directly in the machine or over DHCP. However, cloud providers generally opt to use NAT~\cite{holdrege_ip_1999} to route public IP addresses to the private IPs of servers. This has multiple benefits, including flexibility (public IPs can be changed dynamically without host involvement), security (tenants cannot spoof IPs), and ease of management (centralized view of IP address allocations).

\begin{figure}
    \centering
    \includegraphics[width=\columnwidth]{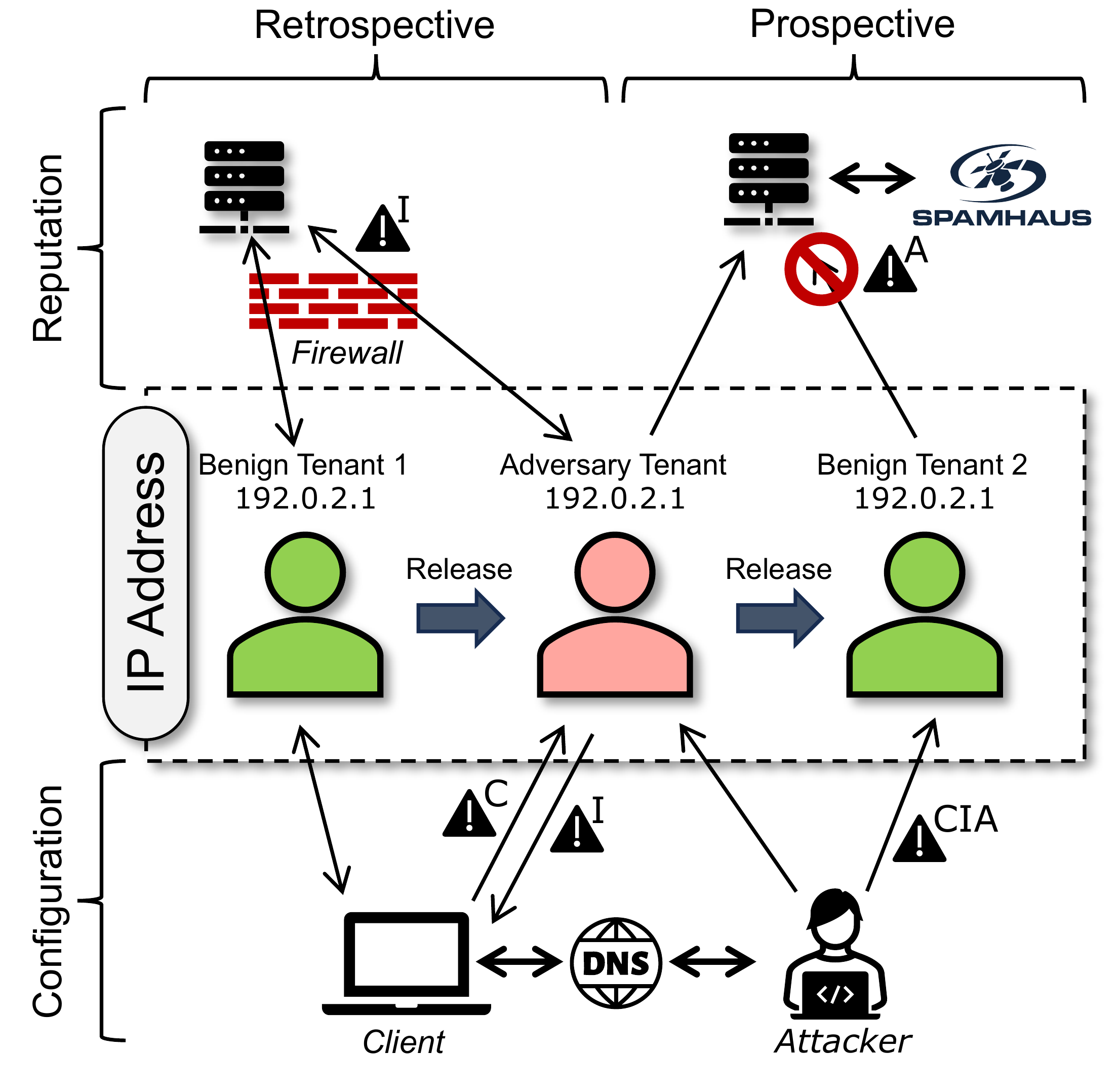}
    \caption{Taxonomy of threats (\includegraphics[height=10pt]{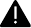}) to the (\texttt{C})onfidentiality, (\texttt{I})ntegrity, and (\texttt{A})vailability of cloud-based network services from IP address reuse. Threats apply to previous tenants (retrospective), future tenants (prospective), and leverage the reputation of IP addresses or associated configuration.
    }
    \label{fig:cloud_squatting}
\end{figure}

\paragraph{Cloud Provider IP Allocation.}
When a tenant requests an IP address, cloud providers have a choice to return any unused address they control, subject to their own internal policy. For instance, a recent work~\cite{pauley_measuring_2022} showed that Amazon Web Services samples their pool of available addresses pseudo-randomly subject to a 30-minute delay between reusing any given address. Another study~\cite{almohri_predictability_2020} found that IP reuse followed a random process, though the ranges of used IP addresses could be inferred from many samples of the pool. Other works have found that Microsoft Azure~\cite{borgolte_cloud_2018} and Google Cloud Platform~\cite{almohri_predictability_2020} show allocation behavior consistent with random allocation. While this random allocation can have the positive effect of allowing for a moving-target defense~\cite{almohri_predictability_2020}, wherein tenants move around the IP address pool to evade attack, it can also lead to severe security weaknesses as discussed below.

\paragraph{The Security Role of IP Addresses.}
When viewed solely as a means to route traffic, IP addresses serve little security role. However, addresses have long been used in the capacity of \textit{security principals}, i.e., control of an IP mediates access to resources, is associated with reputation, and can lead to the receipt of sensitive data. Firewall rules may filter access to specific IP addresses~\cite{ioannidis_implementing_2000}, servers may block messages from historical spam IPs~\cite{spamhaus, antonakakis_building_2010}, and DNS can cause clients to send data to addresses~\cite{liu_all_2016,borgolte_cloud_2018,pauley_measuring_2022}.

\subsection{Exploiting IP Address Reuse}

Due to the use of IP addresses as security principals, the (necessary) reuse of IPv4 addresses by cloud providers opens a set of vulnerabilities to attackers~\cite{ioannidis_implementing_2000, sinha_shades_2008, zhang_highly_2008,antonakakis_building_2010,liu_all_2016, borgolte_cloud_2018, pauley_measuring_2022}. Depicted in \autoref{fig:cloud_squatting}, these vulnerabilities allow adversaries to compromise the confidentiality, integrity, and availability guarantees of the network to other tenants in a variety of ways. We taxonomize such vulnerabilities into those that affect previous tenants (retrospective) and those that affect future tenants (prospective). Further, vulnerabilities may be related to the reputation of the IP address (and associated accessibility of other network services) or to configuration associated with that IP (and associated inbound traffic). Described below, these threat scenarios present different avenues for exploitation, though all rely on the ability for adversaries to acquire and route traffic over a sampling of cloud IP addresses.

\pagebreak[4]
\paragraph{Reputation Attacks.} Source IP address is used to mediate access to a variety of resources on the public Internet. When an adversary uses an address to abuse other services (e.g., by sending spam email, malicious traffic, or large request volume) services may respond by blocking the address~\cite{ioannidis_implementing_2000, sinha_shades_2008, zhang_highly_2008} and reporting to centralized reputation services (e.g., Spamhaus for email~\cite{spamhaus}). This poses a prospective threat to network availability for future tenants. When a future tenant attempts to access services, their address may be blocked because of the actions of previous tenants. Cloud providers pay careful attention to the reputation of their IP pools for services such as managed email sending~\cite{esquivel_effectiveness_2010}, and routinely pay services to \textit{clean} the reputation of their address space. The reputation of IP address ranges is also a key factor in the sale of address blocks~\cite{howard_networking_2020}. Indeed, it is clear that prospective reputation threats to future tenants are a widespread and important issue, though one that to-date has seen little attention in terms of affecting address allocation.

Reputation can also pose risks retrospectively, though such attacks have not yet been observed in practice. Consider a service that mediates access via IP address allowlists~\cite{ioannidis_implementing_2000}. A benign tenant may be granted access to restricted systems via their cloud-allocated IP, and then later release the IP address to the pool. An adversary can then allocate the IP address, and have access to the restricted service via the firewall allow rule. This attack is exceedingly difficult to perform, as the adversary must acquire the IP through random sampling then \textit{also} determine additional services that may be accessible. However, if a service is known to authorize access to a variety of customers via their IP address, it may be possible to quickly enumerate a cloud provider's address space and discover authorized IPs.

\paragraph{Configuration Attacks.} Tenants use IP addresses to refer to resources hosted on cloud providers, causing clients to connect to the resources and establishing trust relationships. Recent works have shown that, when tenants fail to remove the configurations referring to IP addresses they no longer control, these \textit{latent} configurations can be exploited by future tenants~\cite{liu_all_2016, borgolte_cloud_2018, pauley_measuring_2022}. Clients continue to send sensitive data, which is often unencrypted due to trust in the network isolation of the cloud provider. This \textit{retrospective configuration} vulnerability is relatively easy for adversaries to exploit \textit{en masse} on popular cloud providers, as the rapid and random reuse of IP addresses leaves little time for organizations to correct latent configurations. This leaves a long \textit{window of vulnerability} during which adversaries could identify and exploit latent configuration.  The community has proposed methods for correcting configurations such that they do not become latent, but changes to IP address allocation can also play a role when tenants fail to take action.

\pagebreak
While of lower impact, configuration can also pose risks to services prospectively. Here, a tenant (denoted as adversarial although they may be relatively benign) may host services and create a configuration that causes large volumes of traffic to be sent to the address. This traffic could be sourced from legitimate services or from attackers targeting deployed software with exploits or denial of service attacks due to the services a tenant hosts. After releasing the IP, it is allocated to a new (benign) tenant, which then receives the malicious or high-volume traffic targeted at the previous tenant. At a minimum, such high-volume traffic can impose a cost on the new tenant, since cloud providers still charge for outbound bandwidth due to unwanted requests.

\subsection{Preventing exploitation of IP Reuse.} A commonality of all the above attacks is that they rely on the adversary allocating a vulnerable IP address. While the random nature of IP address allocation ostensibly makes the attacks untargeted, prior works have shown that adversaries can easily allocate thousands or even millions of addresses. Because allocation by major providers is currently pseudorandom~\cite{borgolte_cloud_2018, pauley_measuring_2022}, vulnerabilities spanning many IP addresses become akin to the \textit{birthday paradox}, wherein the probability of some adversary IP overlapping with some vulnerable IP quickly approaches 100\%.

Changes to IP allocation policies have been shown to reduce the exploitability of IP Reuse. The goal here is to both (a) reduce the number of IPs that an adversary can allocate, (b) reduce the number of vulnerable tenants associated with those IPs, and (c) increase the window of time between reuse such that associated factors (configuration and reputation) have time to decay. While initial techniques towards achieving this have been proposed~\cite{borgolte_cloud_2018, pauley_measuring_2022}, the community's understanding of the space of attacks and countermeasures here remains incomplete: that is, the ways in which an adversary might adapt to new techniques have not yet been modeled, and resulting further improvements to IP allocation strategies have yet to be explored. Hence, such important questions are the key focus of our work.

\section{Modeling the IP Address Pool}\label{sec:methodology}

Here, we present a comprehensive, novel framework for modeling secure IP address allocation. Towards this, we propose statistical models for tenant behavior (resource allocation and latent configuration), describe algorithms for allocation policies (including our proposed \textit{IP Scan Segmentation} policy), and define threat models under which adversaries might exploit cloud resources. In each case, our methodology is informed by prior works, and validated based on real-world allocation and configuration datasets. \textit{Note: } a reference of symbols used throughout the paper can be found in \autoref{sec:symbols}.

\pagebreak
\subsection{Tenant Behavior} \label{sec:methodology:behavior}
Cloud providers lease resources (e.g., IPs) to tenants under two general paradigms: static and dynamic~\cite{chaisiri2011cost,hwang2013cost,calheiros2011virtual,wolke2015more,bhavani2014resource}. Static allocation allows tenants to acquire a specified amount of resources (perhaps for a fixed period of time); such resources are often used to handle workloads with known or predictable behavior. On the other hand, dynamic allocation allows tenants to acquire and release resources on-demand (to specified upper and lower limits); such resources are typically backed by auto-scalers and other automation tools to handle less predictable workloads efficiently~\cite{qu2018auto}. As such, we model the behavior of tenants within a spectrum of potential allocation strategies (defined in terms of the number of IPs currently allocated to the tenant) spanning static and dynamic resource allocation.

Benign tenants independently allocate IP addresses at some time $t_a$ from the pool and release those addresses at a later time $t_r > t_a$ (here, the IP is said to be allocated for $d_{a}=t_r-t_a$). Tenants also associate configuration with IP addresses, which is dissociated from the IP at $t_c$. Each tenant's overall behavior $B_i$ with respect to IP allocation can therefore be described as a set of timestamps: $$ B_i = \{(t_{a,0},t_{r,0},t_{c,0}), ...,(t_{a,n},t_{r,n},t_{c,n})\}, $$ where $n$ is the total number of IPs allocated to the tenant. A single tenant's behavior then has a maximum limit of $S_{max}$ servers and minimum limit of $S_{min}$ servers; this can capture both static ($S_{max}=S_{min}$) and dynamic ($S_{max}>S_{min}$) resource allocation. For the purposes of our experiments, we focus primarily on dynamic allocations using auto-scalers, as we found this to be most representative of cloud tenant workloads~\cite{qu2018auto,yuan_scryer_2017}.

We next model each tenant's behavior as being independently sampled from a distribution of potential tenant behaviors: $B_i \sim \mathcal{B}$. We approximate $\mathcal{B}$ as a randomized $n$-term Fourier series with a base period of one day~\cite{yuan_scryer_2017}. The intuition is that a given tenant's resource needs will likely vary throughout the day as demand peaks and subsides, but for a given tenant, this pattern will likely be similar from day to day. One work~\cite{yuan_scryer_2017} suggests modeling with a period of 1 week for more precision. Our framework is flexible in this regard, but simulations are performed with 1-day periods. Recall that, by the Shannon-Nyquist sampling theorem~\cite{shannon1949communication}, \textit{any} daily-periodic function can be approximated by a Fourier series of sufficient terms. We compute the tenant's server utilization as a function of the current time $t$ ($0\le t\le 1$), where $0$ and $1$ represent the beginning and end of the day, respectively. We then model the mean server usage of the tenant ($\Bar{S}=\frac{S_{max}+S_{min}}{2}$) and the relative deviation from the mean server usage using the Fourier series:

\begin{equation*}
    \dfrac{S(t)-\Bar{S}}{S_{max}-S_{min}} = \dfrac{\sum_{i=1}^{n} \frac{a_i}{i} \sin(2\pi i (t+\phi_i))}{\sum_{i=1}^{n} \frac{a_i}{i}},
\end{equation*}

\noindent where the Fourier amplitudes ($a_i$) and phases ($\phi_i$) are randomly sampled from the range $[0,1]$. This series has an expected range of $[-0.5,0.5]$, spanning from $S_{min}$ to $S_{max}$ throughout a simulated day. The tenant then allocates or releases IP addresses to respond to this change in compute needs~\cite{herbst2013elasticity}. In keeping with the behavior of a major cloud provider~\cite{aws_autoscale_termination}, IP addresses allocated under autoscale behavior are selected at random for release when a tenant scales down infrastructure.

Modeling autoscaling behavior as a Fourier series creates traces of tenant allocation that are sufficiently realistic to simulate allocation policies. However, on its own, it fails to account for the fact that IP allocations in a given cloud provider region would likely be correlated (due to the local geographies served by that region~\cite{he2013next}). We account for this by biasing the sampling of the lowest-frequency phase of the Fourier series ($\phi_1$): enforcing that $\phi_1 < 0.5$, for instance, will roughly align peak loads to one half of the day. Moreover, tenants may have multiple workloads deployed under the same account that exhibit a hybrid of the above and other behaviors. While evaluation of these hybrid allocation behaviors is beyond the scope of this work, we note that \toolname can also be extended to support other models (or distributions) of tenant behavior, as well as real-world allocation traces. Analysis on real allocations (\autoref{sec:technique_eval:real}) further support findings based on Fourier-distributed allocations, though effectiveness could vary on other workloads.

\subsection{Latent Configuration}

As discussed above, tenants associate configuration with IP addresses when they are allocated. In most cases, this configuration is dissociated from the IP when or before the IP is released ($t_c \le t_r$). In some cases, however, the configuration remains ($t_c > t_r$). If an adversary manages to allocate the IP address before $t_c$, we consider the adversary to have exploited the configuration. The time between IP release and latent configuration ($t_c-t_r$) is the \textit{duration of vulnerability} $d_{v}$ for a given tenant and IP. 

Tenant behavior in dissociating configuration can be highly diverse. For feasibility, we model this configuration dissociation as a Poisson process. We assume that with some probability ($p_c$, a simulation parameter) the tenant leaves latent configuration. If latent configuration is left, it will be dissociated from the IP after some duration $d_v=t_c-t_r$. We model this as an exponential distribution $$d_{v} \sim \textrm{Exponential}(1/d_{a}),$$ where the duration of vulnerability is distributed proportionally to the duration of allocation. Recall the probability density function of such a distribution:

\begin{equation*}
    f(d_{v}) = \begin{cases}
    \frac{1}{d_{a}} e^{-\frac{d_{v}}{d_{a}}}&d_{v}\ge 0 \\
0&d_v<0\\
\end{cases}.
\end{equation*}

This distribution approximates the relationship between the duration of vulnerability and duration of allocation. It reflects empirical observations of cloud deployments~\cite{tang2015holistic}, where relatively short-lived allocations are often orchestrated by automation tools and receive frequent configuration updates (and thus are less prone to having latent configurations), and relatively long-lived allocations are often configured manually and receive infrequent configuration updates (and thus are more prone to having latent configurations). In analysis of data on real-world latent configuration (\autoref{sec:technique_eval:lc}), we find additional evidence supporting this distribution.

\subsection{Adversarial Behavior}

Within a public cloud, an adversary aims to obtain a large number of IP addresses with the goal of exploiting IP reputation, trust, or configurations by previous tenants. We proceed by describing the threat model and capabilities of such adversaries, followed by two modes of behavior: single-tenant (proposed by a prior work~\cite{pauley_measuring_2022}) and multi-tenant (a new consideration of this work).

\subsubsection{Threat Model}
Our work considers an adversary attempting to scan a cloud provider's IP address pool to exploit both address reputation~\cite{ioannidis_implementing_2000, sinha_shades_2008, zhang_highly_2008} and latent configuration left by other tenants (as demonstrated in ~\cite{borgolte_cloud_2018, liu_all_2016, pauley_measuring_2022}). This adversary has no privileged access to cloud resources, and bypasses no security controls in place. Instead, they can only provision resources using paid cloud accounts on a platform. In addition, the adversary can perform a Sybil attack, wherein they control a large number of cloud accounts that are indistinguishable from unique paid customers (\eg by stealing credentials from other accounts, a common attack vector~\cite{gardner_how_2023, goodin_developers_2023, page_microsoft_2024}). We parameterize adversaries by their compute budget (in unique IPs allocated simultaneously) and number of cloud accounts. While these may not be a direct financial cost to an adversary who steals accounts or payment details, they do still represent an opportunity cost, as these credentials could be used for other profitable purposes. The goal of this work is to decrease the effectiveness and increase the cost of such an attack as much as possible.

Within our scenario, the adversary has the capability to allocate IP addresses through public cloud offerings (\eg Amazon EC2). Because we assume the cloud provider cannot soundly determine which tenants are controlled by the adversary, it must serve all tenant requests that are within policy.\footnote{Under a more relaxed threat model, the cloud provider may refuse to service allocations from accounts that are likely malicious. Such actions are fully compatible with and complementary to our approach.} For instance, allocating many instances and IP addresses is commonly used for autoscaling and short-lived tasks\cite{yuan_scryer_2017}. A cloud provider's actions must be a subset of those that would occur under existing offerings. For instance, while a cloud provider must allocate IPs to paying tenants, it may choose any free IP address to allocate. Based on this threat model, prior works~\cite{borgolte_cloud_2018,pauley_measuring_2022} have proposed a single-tenant adversary that allocates IPs under one tenant. This work considers a stronger adversary that has access to multiple tenants, defeating existing defenses through a Sybil attack.

\subsubsection{Single-tenant Adversary}\label{sec:methodology:stadv} Discussed in prior works ~\cite{liu_all_2016, borgolte_cloud_2018, pauley_measuring_2022}, a single-tenant adversary provisions IP addresses under a cloud account with the aim of allocating many unique addresses. In most cases, the most effective means by which to do this is to rent virtual servers with an associated IP address. A tenant allocates many of these servers simultaneously, runs them for the minimum time required to discover vulnerable configuration or leverage the address reputation, and then releases the IPs back to the provider (or retains the server if there is interesting configuration associated). In this way, the tenant can easily sample from the IP address space unless the provider takes steps to prevent it. In line with cloud provider service quotas on concurrent allocations~\cite{aws_quotas}, our simulated single-tenant adversary allocates up to 60 IPs simultaneously for 10 minutes each, before releasing the IPs and allocating new ones.

\subsubsection{Multi-tenant Adversary}\label{sec:methodology:mtadv} The multi-tenant adversary adapts to protective allocation policies by leveraging multiple tenants for allocations. An adversary could create multiple tenants using Virtual Private Networks and private credit cards to evade detection\footnote{Note that our threat model assumes the adversary is still cost-limited, either directly or in ability to acquire usable stolen credit card numbers.}. Under this threat model, we also assume that a cloud provider must make allocation decisions based solely on tenant behavior, and cannot identify collusion between tenants otherwise. Further, the cloud provider must prioritize availability, and so must grant tenant allocation requests even if they believe the tenant to be malicious. Due to these factors, the multi-tenant adversary represents a stronger threat model that existing allocation policies may not protect against. In the worst case (and as simulated in \autoref{sec:technique_eval:mtadv}), the adversary would continually use new accounts after allocating the maximum concurrent IPs on a single account.

\subsection{IP Allocation Policies}
When tenants request an IP address from the cloud provider, the provider can choose which IP to assign to the tenant. Here, we assume (and prior works have shown~\cite{liu_all_2016}) that the cloud provider can freely choose to assign any free IP address within some zone to a tenant, and that there is no technical restriction on when IPs get reused. As noted (\autoref{sec:background:allocation}), cloud providers use NAT to route public IP addresses, so assignment of these addresses can happen instantaneously and without any restriction from the underlying network topology.

Within this framework, the policy is a stateful set of functions that \textsc{Allocate}, \textsc{Release}, and \textsc{Init} IP addresses:

\paragraph{$\text{\textsc{Allocate}}(T,\theta)\longrightarrow (ip,\theta')$}: Accepts a tenant id $T$ and an opaque state $\theta$ (for tracking IP allocation parameters) and returns a new, usable IP for the tenant, as well as an updated opaque state $\theta'$.

\paragraph{$\text{\textsc{Release}}(ip,\theta)\longrightarrow (\theta')$}: Accepts an allocated $ip$ (previously allocated by some tenant id $T$) and an opaque state $\theta$ and releases the IP back to the pool, returning an updated opaque state $\theta'$.

\paragraph{$\text{\textsc{Init}}(ip,\theta)\longrightarrow (\theta')$}: Accepts a new $ip$ into the pool that was never previously allocated, and returns an updated state $\theta'$.
\vspace{6pt}

All calls to \textsc{Allocate} and \textsc{Release} are paired in order, such that IP addresses are in use by at most one tenant at a time. 

We next describe different allocation policies considered in our evaluation, providing their implementation in natural language and pseudocode. Note that if the \textsc{Release} and \textsc{Init} interfaces are not provided, it is assumed that the default implementations presented in \autoref{alg:releaseinit} are used. Of these policies, the \Random policy is implemented in practice by cloud providers~\cite{pauley_measuring_2022, borgolte_cloud_2018}, and the \LRU and \Tagged policies were proposed by a prior work~\cite{pauley_measuring_2022}. In addition to these policies that encompass the current state of the art, we propose and evaluate a new policy, \textit{IP scan segmentation}.

\begin{betteralgorithm}[t]
\setstretch{1.1}
\Fn{\REL($ip, \theta$)}{
    $ip.t_r \gets \getCurrentTime()$\;
    $\theta' \gets \setIpNotAllocated(\theta, ip)$\;
    \Ret{$\theta'$} 
}
\Fn{\INIT($ip, \theta$)}{
    $\theta' \gets \createIp(\theta, ip)$\;
    \Ret{$\theta'$} 
}
\caption{Default \textsc{Release} and \textsc{Init} interfaces }
\label{alg:releaseinit}
\end{betteralgorithm}

\subsubsection{Pseudorandom (\Random, \autoref{alg:pr}).}
The most basic IP allocation policy (and that used by major cloud providers~\cite{almohri_predictability_2020,pauley_measuring_2022}) is pseudorandom allocation. Here, IPs are sampled randomly from the pool of available addresses, with the only restriction being that IPs cannot be used within $d_{reuse}$ (\SI{30}{min} as observed by prior work~\cite{pauley_measuring_2022}). It has benefits for ease of use and understanding, as minimal information needs to be associated with the address. Further, the pool could be managed in a distributed fashion (such as within separate datacenters).

\begin{betteralgorithm}[t]
\setstretch{1.1}
\Fn{\PRALL($T, \theta$)}{
    $ip \gets \randomSample(\mathcal{I} \setminus \mathcal{I}_{A_{t}})$\;
    \While{$ \getCurrentTime() - ip.t_r < d_{reuse}$}{
        $ip \gets \randomSample(\mathcal{I} \setminus \mathcal{I}_{A_{t}})$\;
    }
    $\theta' \gets \setIpAllocated(\theta, ip)$\;
    \Ret{$ip, \theta'$} 
}
\caption{(\Random) IP Allocation}
\label{alg:pr}
\end{betteralgorithm}

\subsubsection{Least Recently Used (\LRU, \autoref{alg:lru}).} The LRU policy seeks to maximize the median time between reuse of IP addresses. It does this by always allocating the IP address that has been in the pool the longest. Such an algorithm can either be implemented deterministically (\eg using a FIFO queue), or stochastically (\eg by sampling a subset of the IPs in the pool and returning the oldest of that batch). Such stochastic approaches have been shown to achieve acceptable performance in practice for caches~\cite{sanfilippo_random_nodate}.

\begin{betteralgorithm}[t]
\setstretch{1.1}
\Fn{\LRUALL($T, \theta$)}{
    $ip \gets \underset{ip \in \mathcal{I} \setminus \mathcal{I}_{A_{t}}}{\argmin(}ip.t_r)$\;
    $\theta' \gets \setIpAllocated(\theta, ip)$\;
    \Ret{$ip, \theta'$} 
}
\caption{\LRU IP Allocation}
\label{alg:lru}
\end{betteralgorithm}

\subsubsection{IP Tagging (\Tagged, \autoref{alg:tag}).}
Recent work~\cite{pauley_measuring_2022} presented the first IP allocation policy specifically intended to prevent adversaries from scanning the IP pool. Referred to as \textit{IP Tagging}, the authors describe that, intuitively, released IP addresses are tagged with the tenant ID that released them. When allocating an IP, tenants first preference the IP addresses that they are tagged to, followed by addresses tagged to any other tenant using LRU allocation. Our implementation additionally stipulates that tagged IP addresses are selected in an LRU fashion, though other variants such as selecting the most-recently-used tagged IP may also be valid approaches. In any case, selecting a tagged IP address inherently exposes no additional IP address or tenant configuration to an adversary. Our evaluation also further characterizes IP Tagging beyond the metrics performed in prior work to assess the generality of the technique to stronger adversaries.

\begin{betteralgorithm}[t]
\setstretch{1.1}
\Fn{\IPTAGGINGALL($T, \theta$)}{
    \eIf{$ \exists ip \in \mathcal{I} \setminus \mathcal{I}_{A_{t}} \mid ip.ID = T$}{
        $ip \gets \underset{ip \in \mathcal{I} \setminus \mathcal{I}_{A_{t}}}{\argmin}(ip.t_r \mid ip.ID=T)$\;
    }{
        $ip, \_ \gets \LRUALL(T, \theta)$\;
    }
    $ip.ID \gets T$\;
    $\theta' \gets \setIpAllocated(\theta, ip)$\;
    \Ret{$ip, \theta'$} 
}
\caption{\Tagged IP Allocation}
\label{alg:tag}
\end{betteralgorithm}

\subsubsection{IP Scan Segmentation (\autoref{alg:segmentation}).}
While IP tagging provides protection against a single-tenant adversary, the technique could be susceptible if an adversary spreads allocations across many tenants, bypassing the tagging entirely. In response to this threat, and our more powerful characterization of pool scanning adversaries (\autoref{sec:methodology:mtadv}), we propose a new IP allocation policy that aims to prevent IP scanning by adversaries even when the adversary has access to an arbitrary number of cloud tenants.

\begin{figure*}
    \centering
    \includegraphics[width=\linewidth]{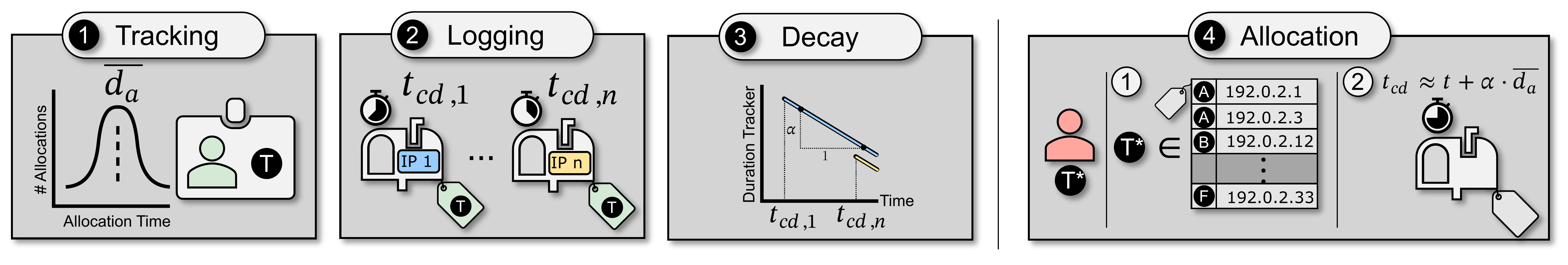}
    \caption{IP Scan Segmentation - \ding{202} The mean IP allocation duration for tenant $T$ is tracked (i.e., $\Bar{d_a}$), \ding{203} each released IP $n$ is first associated (i.e., tagged) with tenant $T$ \& the allocated duration , \ding{204} the duration associated with the IP $n$ then decays linearly with rate $1/\alpha$ (stored as the cooldown time $t_{cd}$), and \ding{205} when an IP is allocated for tenant $T^*$, preference is first given to a  $T^*$-tagged IP, then to an IP from the general pool whose $t_{cd}$ is closest to $t+\alpha\cdot\Bar{d_a}$.
    }
    \label{fig:segmentation_explained}
\end{figure*}

Our proposed policy, \textit{IP scan segmentation} (shown in \autoref{fig:segmentation_explained}), works by identifying tenant behavior that is indicative of (and necessary for) IP pool scanning. The pool tracks the \textit{mean allocation time} ($\Bar{d_a}$) for each tenant $T$: relatively long-lived resources will lead to high $\Bar{d_a}$, and adversarial scanning (which inherently must allocate many IPs) would require a low $\Bar{d_a}$ to be economically feasible. IP addresses are tagged with both (a) the ID of the most recent tenant, %
and (b) the duration the IP was allocated for (this decays over time, see \textit{cooldown time}). If the IP was previously held for longer, this value does not change (so that a short allocation does not delete the protection from a previous longer allocation).

When a tenant allocates an IP address, preference is first given to an IP tagged to that tenant (as in IP Tagging), followed by an IP from the pool that was previously allocated for as close as possible to $\Bar{d_a}$. In this way, adversary tenants that scan the IP space will in turn be allocated IP addresses that were previously allocated for short periods of time, either by another adversary tenant or by tenants deploying short-lived workloads (which are less likely to have associated latent configuration).

\paragraph{Cooldown time.} As noted above, each IP is tagged with the longest duration it has been held for. Over time, this approach alone would cause more and more IPs to be tagged with long duration, leaving fewer and fewer with short durations and eventually allowing scanners to allocate the IPs that should be protected. Due to the scarcity of IP addresses, granting every IP address high protection means no IP receives protection.

To prevent this, the \Segmented policy applies a cooldown to the allocation duration over time with rate $1/\alpha$. The duration associated with an IP is therefore $d_a-(t-t_r)/\alpha$. Rather than continually update this duration, the \Segmented policy tracks the x-intercept of this function. This intercept, the \textit{cooldown time} of the IP, is the time when the IP will no longer be provided any protection by the \Segmented policy. To select the IP with the most similar allocation duration for a given tenant, the policy minimizes $|(t_{cd}-t)-\alpha\cdot\Bar{d_a}|$. In this way, tenants receive IP addresses that have exhibited similar allocation behavior to their past allocation behavior. Additionally, new tenants start with $\Bar{d_a}=0$, so they will receive IPs that have been segmented for allocation to scanners.

Since adversarial scanning would require a low $\Bar{d_a}$ to be economical, an adversary tenant would then be matched with IPs that were either released a long time ago or were kept for a very short amount of time, mitigating some of the risk of an adversary acquiring an IP with latent configuration. Further, IPs recently released by the adversary would have a $t_{cd}$ consistent with their average allocation duration, increasing the likelihood that they receive the same IP back even under a different tenant.

\begin{betteralgorithm}[t]
\setstretch{1.1}
\Fn{\IPSCANSEGMENTATIONALL($T, \theta$)}{
    $T.n_a \gets T.n_a + 1$\;  
    \eIf{$ \exists ip \in \mathcal{I} \setminus \mathcal{I}_{A_{t}} \mid ip.ID = T$}{
        $ip \gets \underset{ip \in \mathcal{I} \setminus \mathcal{I}_{A_{t}}}{\argmin}(ip.t_r \mid ip.ID=T)$\;
    }{
         $ip \gets \underset{ip \in \mathcal{I} \setminus \mathcal{I}_{A_{t}}}{\argmin(}|ip.t_{cd}-\getCurrentTime()-\alpha\cdot T.\Bar{d_a}|)$\;
    }
    $ip.ID \gets T$\;
    $ip.t_a \gets \getCurrentTime()$\;
    $\theta' \gets \setIpAllocated(\theta, ip)$\;
    \Ret{$ip, \theta'$} 
}
\Fn{\IPSCANSEGMENTATIONREL($ip, \theta$)}{
    $ip.t_r \gets \getCurrentTime()$\;
    $ip.t_{cd} \gets ip.t_r + \alpha\cdot(ip.t_r-ip.t_a)$\;
    $T \gets ip.ID$\;
    $T.d_a \gets T.d_a + ip.t_r - ip.t_a $\;
    $\theta' \gets \setIpNotAllocated(\theta, ip)$\;
    \Ret{$\theta'$} 
}
\caption{\Segmented IP Allocation}
\label{alg:segmentation}
\end{betteralgorithm}

\begin{figure}[t]
    \centering
    \includegraphics[width=\columnwidth]{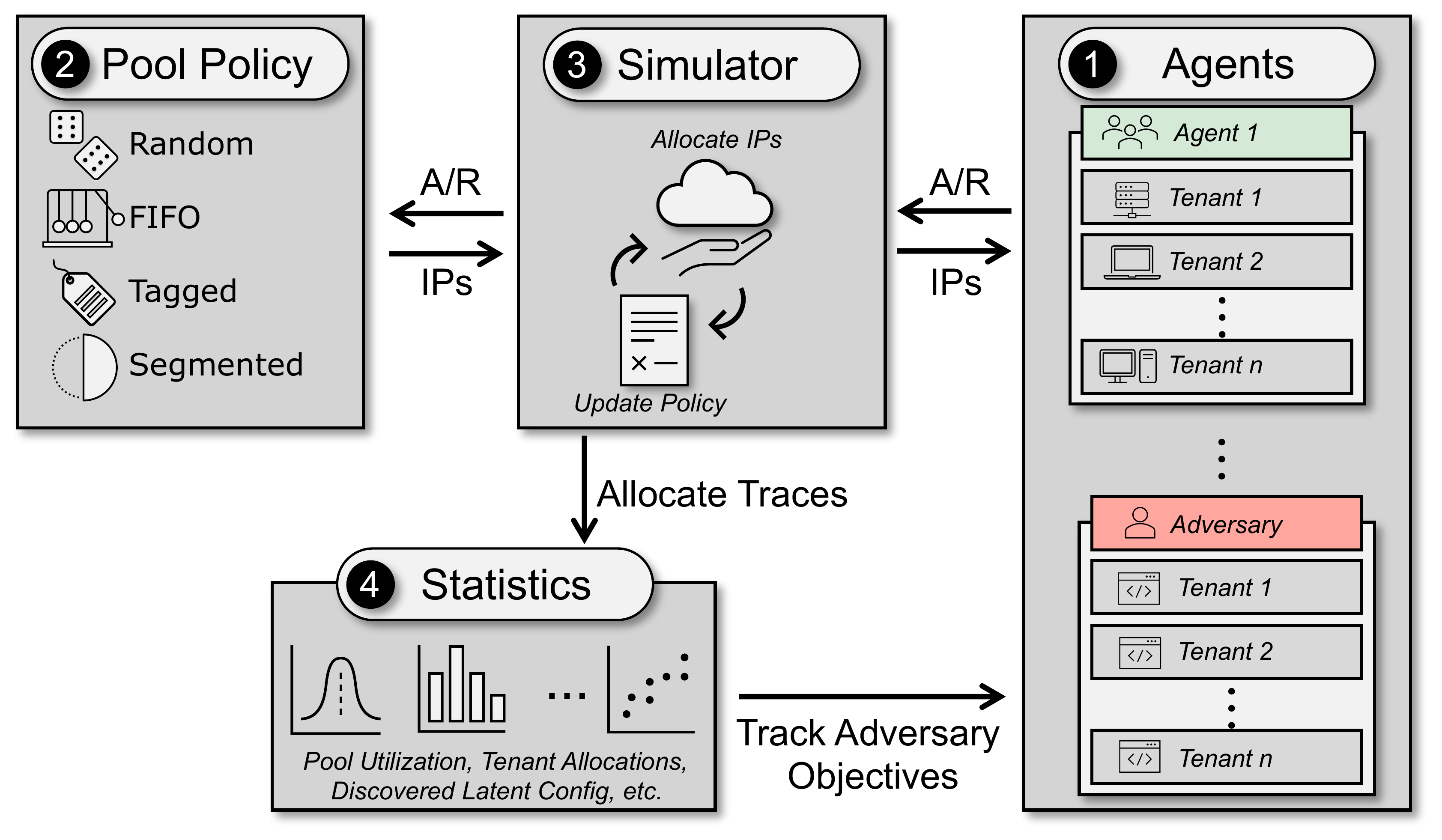}
    \caption{Overview of our analysis - \ding{202} We first define agents who (A)llocate and (R)elease IP addresses in varying modalities (including adversarial behaviors), \ding{203} we then evaluate a suite of IP pool allocation policies that govern IPs associated with tenants, \ding{204} we then simulate interactions between agents and policies, and \ding{205} collect various statistics concerning pool utilization, adversarial goals, etc.}
    \label{fig:overview}
\end{figure}

\subsection{Implementation}

To empirically study the distribution of IP allocation behaviors, adversarial techniques, and cloud provider defenses, we develop an IP pool simulator (\toolname). \toolname implements an extensible and configurable architecture for simulating interactions with IP address pools. Described in \autoref{fig:overview} and at \href{https://github.com/MadSP-McDaniel/eipsim/}{GitHub.com/MadSP-McDaniel/eipsim/}, \toolname allows us to evaluate the efficacy of allocation policies across a variety of scenarios over hundreds of years of simulated cloud provider allocation.

\pagebreak
\section{Evaluation}\label{sec:technique_eval}

We proceed by evaluating the security properties of IP allocation policies, first in synthetic settings (this section) followed by evaluations of realism and practicality (\autoref{sec:realism}). We begin by defining simulation parameters and defender objectives, followed by comparison of policies in both benign and adversarial settings.\footnote{Our main analysis considers a multi-tenant adversary. For completeness, we also include evaluations on single-tenant adversaries (those considered in prior works) in \autoref{sec:technique_eval:stadv}.} Our analysis highlights the positive impact of new allocation policies: in both synthetic and real-world traces, IP Scan Segmentation reduces exploitable latent configuration and IP address yield compared to existing techniques, even when a strong adversary can use many cloud tenants.

\subsection{Simulation Parameters and Objectives}

Our evaluation aims to quantify the impact of environmental, policy, and adversarial conditions on security properties. As such, we perform sweeps of multiple parameters that affect allocation policy performance. Within this setting, a cloud provider (defender) aims to reduce the ability of adversaries to exploit address reputation and previous tenants, objectives that are further defined here.

\paragraph{IP Count and Utilization.} The size of the overall IP pool ($|\mathcal{I}|$), and the number of IPs allocated at any given time $|\mathcal{I}_{A_t}|$) has a substantial impact on allocation performance. If the majority of IP addresses are assigned, for instance, the pool policy has fewer choices when a tenant requests a new IP, and strategies that age, tag, or segment the addresses will therefore be less effective. Here, we can study performance by varying the max pool \textit{allocation ratio} ($AR_{max} = \max_{t} \frac{|\mathcal{I}_{A_t}|}{|\mathcal{I}|}$) between simulations. Our evaluated simulation scenarios have $\max_{t}|\mathcal{I}_{A_t}| \approx \SI{680}{k}$, and compute $|\mathcal{I}|$ using $AR_{max}$ (set in each experiment).

\paragraph{Allocation Duration.} Benign and adversarial tenants allocate IP addresses and hold them for some period of time. Study of the duration for which tenants and adversaries allocate IPs can yield insights on countermeasures. Our simulated adversary holds IPs for 10 minutes.

\paragraph{Free Duration.} Pools hold free addresses available for allocation, and holding an address for longer decreases the likelihood of associated latent configuration. As such, understanding the distribution of how long pools keep IPs free can suggest measures towards reducing latent configuration.

\paragraph{Latent Configuration Probability.} In all simulations, we use a fixed probability of a given tenant leaving latent configuration, $p_c=0.1$. In separate evaluations, we found that results varied roughly linearly with this parameter, making it less interesting for extensive study. However, future works could use more complex models for latent configuration where this constant plays a greater role.

\paragraph{Defender Objectives.} Recall that the goal of the defender is to protect benign tenants against both \textit{retrospective} (i.e., against previous tenants) and \textit{prospective} (i.e., against future tenants) attacks by adversaries:
\begin{itemize}
    \item \textit{Retrospective attacks} are prevented by reducing the amount of \textit{latent configuration} that an adversary can detect per IP allocated (proportional to total cost). We measure this quantity as \textit{latent configuration yield}, the fraction of IP allocations which yield a (1) unique IP address with (2) some associated latent configuration.
    \item \textit{Prospective attacks} are mitigated by shielding future tenants from the negative effects of adversaries. This can be achieved by reducing the number of unique IPs that adversaries can obtain. We therefore also measure \textit{unique IP yield}, which is the fraction of IP allocations which yield a new unique IP address. Reducing unique IP yield likewise reduces the number of IPs whose reputation can be harmed by adversaries.
    Prospective attacks can also be evaluated by the rate of future tenant allocations that have been \textit{poisoned} by an adversary. In ancillary evaluations, we found this metric to mirror that of unique IP yield, so we focus on the latter for extensive study.
\end{itemize}

\begin{figure*}[ht]
    \begin{tabular}{cc}
    
    \begin{tabular}{c}
    \begin{subfigure}[t]{.45\textwidth}
        \centering
        \includegraphics[width=\textwidth]{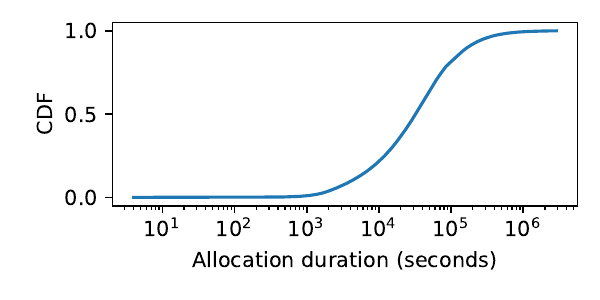}
        \subcaption{Distribution of tenant allocation durations under the scenario described in \autoref{sec:technique_eval:non_adversarial}. The broad range of allocation durations demonstrates the generality of the simulation parameters for evaluating pool policies.}
        \label{fig:benign:allocation_cdf}
    \end{subfigure}\\
    \begin{subfigure}[t]{.45\textwidth}
         \centering
        \includegraphics[width=\textwidth]{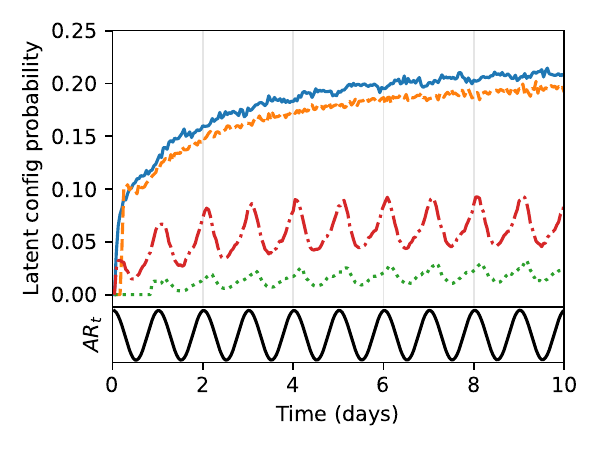}
        \subcaption{Latent configuration prevalence over time ($AR_{max}\approx0.97$). Here, the lower plot shows the instantaneous allocation ratio of the pool over time ($AR_{t}$). As the pool reaches max allocation, strategies tend to allocated IPs with more latent configuration as addresses must be reused more quickly.}
         \label{fig:benign:lc_vs_time}
    \end{subfigure}\\
    
    \end{tabular}\hfill &
    
    \begin{tabular}{c}
    \begin{subfigure}[t]{.45\textwidth}
        \centering
    \includegraphics[width=\textwidth]{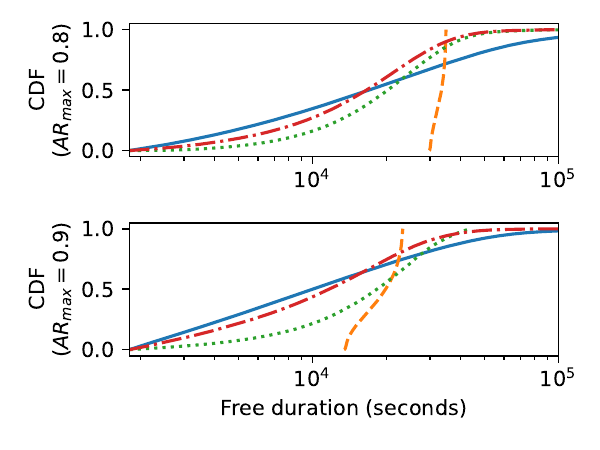}
    \subcaption{Distribution of time between reuse across policies for two values of $RA_{max}$. The \LRU pool sees the most impact from having more IPs available, as addresses can be aged for longer.}
    \label{fig:benign:free_duration_cdf}
    \end{subfigure}
    \\
    \begin{subfigure}[t]{.45\textwidth}
        \centering
        \includegraphics[width=\textwidth]{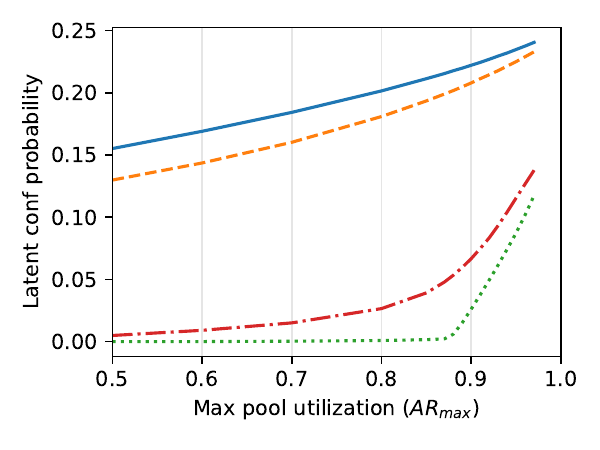}
        \subcaption{Overall latent configuration varying max allocation ratio $AR_{max}$. IP Tagging and Segmentation policies further reduce prevalence even at high allocation ratios.}
        \label{fig:benign:lc_vs_ra}
    \end{subfigure}\\
    \end{tabular}\hfill\\
    \vspace{1pt}
    \end{tabular}
    \centering
    \includegraphics[width=\columnwidth]{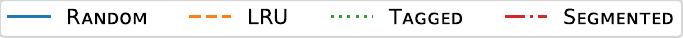}
    \caption{Modeling tenant allocations ($p_c=0.1$).} %
    \label{fig:benign}
\end{figure*}

\subsection{Non-adversarial Scenario}\label{sec:technique_eval:non_adversarial}

To understand the aggregate performance of the various IP allocation policies, we first perform a simulation of the pool with no adversary. Here, agents allocate and release IP addresses on behalf of simulated tenants, and we study the effect of these policies on the configurations associated with allocated addresses. Experiments run for 180 simulated days, with 64 total experiments across allocation ratios and policy parameters (32 years of total simulated allocation).

Results are shown in \autoref{fig:benign}. From these simulation results, we can distill several conclusions about the efficacy of our model and simulator, strengths and weaknesses of existing policies, and insights towards development of new policies.

\pagebreak
\paragraph{Tenant Behavior.} We first analyze the distribution of tenant allocation durations (\autoref{fig:benign:allocation_cdf}). Here, we see that simulated allocations span several orders of magnitude in duration, representing a diverse distribution of behavior. Furthermore, to allocate within the distribution of other tenants an adversary would need to hold IPs and associated servers for an extended period of time, reducing yield for a given cost. This provides hope that adversarial behavior in the pool could be identified and segmented from legitimate users.

\paragraph{Time Between Reuse.} Next, we can see differences in how long policies keep IP addresses between reuse (\autoref{fig:benign:free_duration_cdf}). Results are shown for two allocation ratios ($AR_{max}=0.8$ and $AR_{max}=0.9$). These represent low- and high-contention scenarios for the pool, respectively. Beyond $AR_{max}=0.97$, the policies cannot consistently age IPs for at least 30 minutes before reuse. In both cases, allocation schemes other than \LRU perform similarly, reusing IP addresses in as little as 30 minutes, whereas \LRU consistently maximizes the minimum time between reuse, by design. While this figure implies that \LRU may be superior for preventing latent configuration, other policies that specifically target adversarial allocations may perform better in practice due to other factors.

\begin{figure*}[ht]
    \begin{subfigure}[t]{.32\textwidth}
        \centering
        \includegraphics[width=\textwidth]{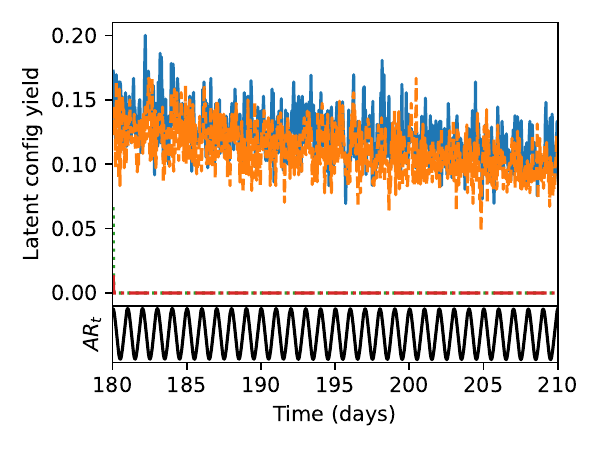}
        \caption{Yield of new latent configurations over time ($AR_{max}=0.9$). The first 10 simulated days are omitted and identical to \autoref{fig:benign:lc_vs_time}.}
        \label{fig:stadv:lc_vs_time}
    \end{subfigure}
    \hfill
    \begin{subfigure}[t]{.32\textwidth}
        \centering
    \includegraphics[width=\textwidth]{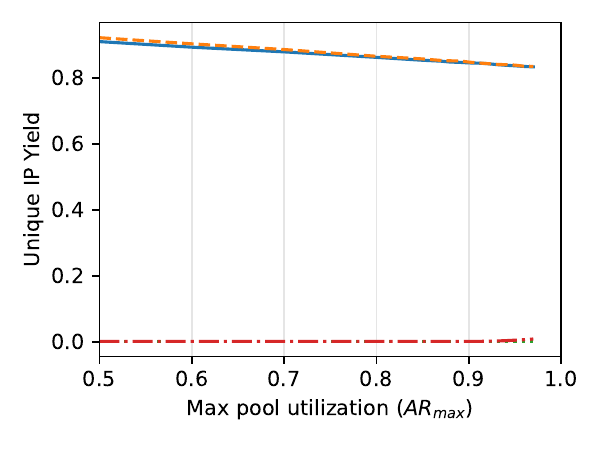}
    \caption{Effect of pool utilization on discovered unique IPs. \Tagged and \Segmented strongly protect against the single-tenant adversary up to high pool utilization.}
    \label{fig:stadv:ips_vs_ra}
    \end{subfigure}
    \hfill
     \begin{subfigure}[t]{.32\textwidth}
         \centering
        \includegraphics[width=\textwidth]{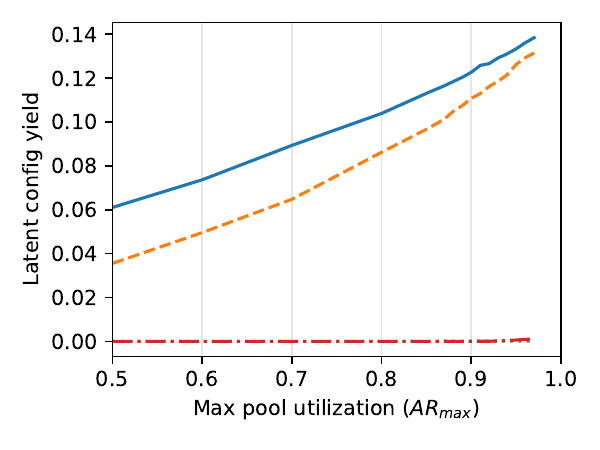}
        \caption{Effect of pool utilization on discovered latent configurations. While \LRU provides mild protection vs \Random, \Tagged and \Segmented have superior performance.}
         \label{fig:stadv:lc_vs_ra}
    \end{subfigure}
    \centering
    \includegraphics[width=\columnwidth]{figs/legend.pdf}
    \caption{Modeling the single-tenant adversary.}
    \label{fig:stadv}
\end{figure*}

\paragraph{Pool Behavior Over Time.} Looking at prevalence of latent configuration over time in \autoref{fig:benign:lc_vs_time}, we initially see lower prevalence as the pool has unused IP addresses to allocate. Beyond that, prevalence for \Random and \LRU allocation approaches $p_c$ (note that prevalence can exceed $p_c$ as multiple tenants have the opportunity to associate configuration with a given IP address). \LRU unsurprisingly outperforms \Random slightly, due to the higher time between reuse of IP addresses. While higher time between reuse most clearly reduces aggregate exposure of latent configuration under our posited exponential distribution, cloud providers could also use \toolname with other models of latent configuration to validate against their unique scenarios. We expect similar results from any monotonic distribution of $d_v$.

\paragraph{Effect of Pool Utilization.} IP addresses are a scarce resource, so cloud providers should aim to achieve the best security against latent configurations while incurring minimal pool size overhead. In \autoref{fig:benign:lc_vs_ra}, we see that the studied allocation policies have differing behavior as pool size changes. While both \Tagged and \Segmented outperform the \Random and \LRU policies, \Tagged performs slightly better. This is because benign tenants preferentially receive IPs from other tenants exhibiting benign behavior, making other IPs available for segmentation to heuristically malicious tenants. Our experiments demonstrate that allocation policies can have marked impact on overall latent configuration exposure even for high IP allocation ratios.

\vspace{6pt}

Our non-adversarial experiments show that \toolname and its associated models are a compelling means by which to study the behavior of IP address pools, spanning a broad range of resulting tenant allocations. The parameters of our initial simulation prove interesting for further study, as the variety of tenant behaviors leads to differentiated performance across allocation policies.

\subsection{Single-Tenant Adversary}
\label{sec:technique_eval:stadv}

We next simulate an adversary that is attempting to explore the IP space and discover latent configurations using only a single account. To do this, we model each simulation as in \autoref{sec:technique_eval:mtadv}, but with a tenant count of one. For each simulated adversary, we seek to answer two questions:

\begin{enumerate}
    \item How many unique IPs can the adversary discover based on their allocation scheme?
    \item How many new latent configurations does the adversary discover associated with those IP addresses?
\end{enumerate}

\paragraph{Unique IPs.} \autoref{fig:stadv:ips_vs_ra} displays the adversary's ability to discover new IPs across policies and allocation ratio. The \Random and \LRU policies exhibit roughly identical behavior: IP yield is reduced as the pool gets smaller ($AR_{max}$ gets higher) because the adversary is more likely to receive the same IPs back. Likewise, \Tagged and \Segmented both almost completely eliminate the single tenant adversary's ability to discover new IPs. This is unsurprising, as both strategies tag IPs to the most recent tenant and reallocate those IPs back to the tenant. \Segmented exhibits a slight increase in adversarial IP yield at very high allocation ratios, as other tenant allocations interfere with the IPs tagged to the adversary--this does not occur in \Tagged because the LRU backup queue prevents the tenant's tagged IPs from being taken.

\paragraph{Latent Configuration.} While an adversary might directly seek to observe a high number of IPs, the end goal is to discover IPs that actually have associated configuration. Our results (\autoref{fig:stadv:lc_vs_ra}) demonstrate a marked difference here as well, with both \Tagged and \Segmented performing equivalently well against the single-tenant adversary. As seen in the non-adversarial scenario, \LRU also slightly outperforms \Random as IP addresses are held in the pool longer before reuse, though this effect is diminished as the allocation ratio increases since the policies are best effort and must allocate some available IP to the adversary.

Our tool also allows us to model adversarial objectives over time (\autoref{fig:stadv:lc_vs_time}). Here, we see that the bulk of latent configuration discovered under \Tagged and \Segmented occurs early in the experiment. Beyond this, the pool returns the same IP addresses to the adversary and no latent configurations are discovered.
\begin{figure*}[!ht]
\begin{tabular}{lr}
    
    \begin{tabular}{c}
    \begin{subfigure}[t]{.45\textwidth}
        \centering
        \includegraphics[width=\textwidth]{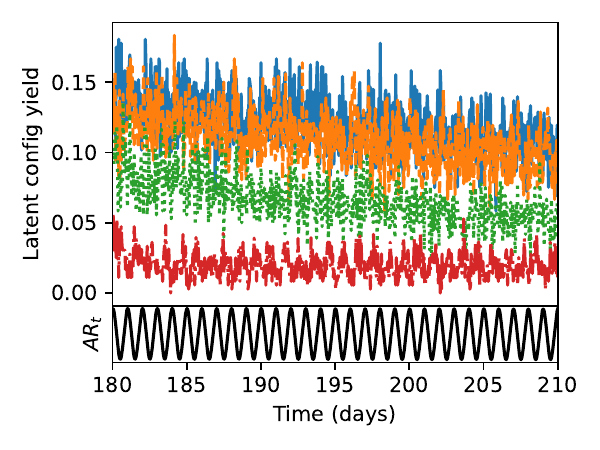}
        \caption{Yield of new latent configurations over time ($AR_{max}=0.9$). Data points are bucketed by hour.}
        \label{fig:mtadv:lc_vs_time}
    \end{subfigure}\\
    \begin{subfigure}[b]{.45\textwidth}
        \centering
    \includegraphics[width=\textwidth]{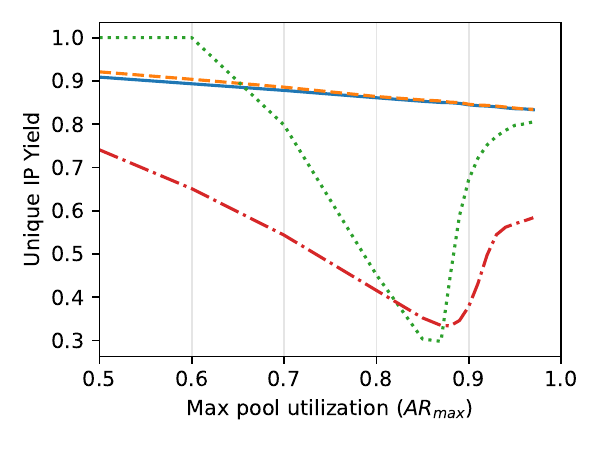}
    \caption{Effect of pool utilization on discovered unique IPs (unlimited adversary tenants)}
    \label{fig:mtadv:ips_vs_ra}
    \end{subfigure}\\
    
    \end{tabular}\hfill &
    
    \begin{tabular}{c}
    \begin{subfigure}[t]{.45\textwidth}
         \centering
        \includegraphics[width=\textwidth]{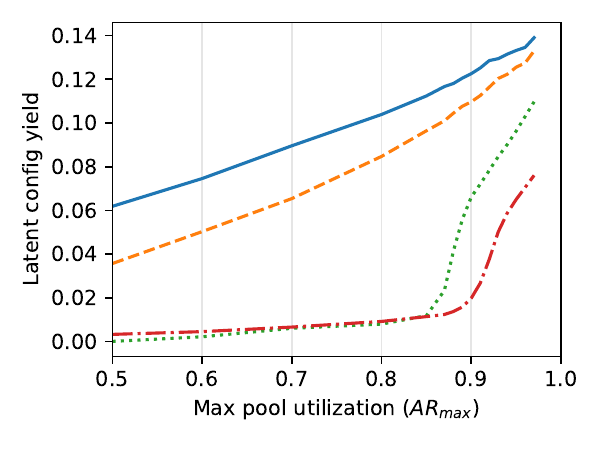}
        \caption{Effect of pool utilization on discovered latent configs (unlimited adversary tenants)}
         \label{fig:mtadv:lc_vs_ra}
    \end{subfigure}\\
    \begin{subfigure}[b]{.45\textwidth}
         \centering
        \includegraphics[width=\textwidth]{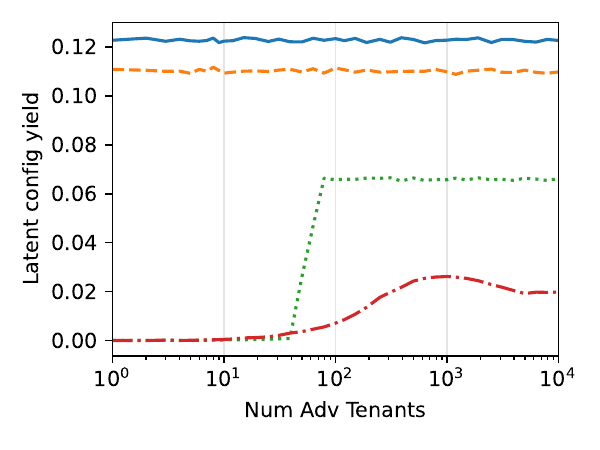}
        \caption{Effect of total adversary tenants on latent configuration yield ($AR_{max}=0.9$)}
         \label{fig:mtadv:lc_vs_tc}
    \end{subfigure}\\
    \end{tabular}\hfill\\
    \end{tabular}
    \centering
    \includegraphics[width=\columnwidth]{figs/legend.pdf}
    \caption{Modeling the multi-tenant adversary.}
    \label{fig:mtadv}
\end{figure*}

\subsection{Multi-tenant Adversary}\label{sec:technique_eval:mtadv}

Next, we evaluate how pool implementations defend against a sophisticated adversary who can use many cloud tenants to bypass the protections of existing allocation policies. We also evaluate how an adversary's success varies with the number of tenants they can create. Simulations are run in the benign setting for 180 days, followed by 30 days of adversarial exploitation. 280 total simulations are performed across allocation ratio, policy, and adversarial tenant count parameters (160 years of total simulated allocation).

\paragraph{Unique IPs.} \autoref{fig:mtadv:ips_vs_ra} shows the number of unique IPs discoverable by an unlimited-tenant adversary as pool utilization varies. While the non-tenant-aware policies \Random and \LRU show no difference from the single-tenant adversary, tenant-aware policies show surprising results. In both cases, unique IPs reduce as utilization increases to some critical point, then increases again. The increased unique IP yield at low utilization results from the large number of free IPs, resulting in increased availability for both benign and malicious tenants. As a result, higher unique IP yield does not result in discovery of exploitable latent configurations. Above the critical point, both \Tagged and \Segmented \textit{must} allocate potentially-dangerous IPs to tenants, but \Segmented successfully identifies behavior patterns across adversary tenants and reduces the number of unique IPs seen. In this way, \Segmented successfully protects a larger portion of the IP space.

\paragraph{Latent Configuration.} \autoref{fig:mtadv:lc_vs_ra} shows how the multi-tenant adversary's yield of latent configuration varies with allocation ratio. Here, we see the complete effect of tenant-aware allocation policies: below the policy's critical point, allocated IPs have minimal associated latent configuration, so a high unique IP yield does not allow exploitation. Above this, strategies offer only mild protection (\ie approaching that of non-tenant-aware policies). Most importantly, however, this plot emphasizes the advantages of IP scan segmentation: \Segmented reduces latent configuration yield by \bestsynmtadvsegmentednewLatentConfsImprovementOverrandom{} compared to \Random, whereas \Tagged only reduces yield by \bestsynmtadvtaggednewLatentConfsImprovementOverrandom{}. In other words, \Segmented achieves an additional reduction of \bestsynmtadvsegmentednewLatentConfsImprovementOvertagged{}. When considering an adversary with the ability to use multiple cloud tenants, \Segmented offers superior protection to prior works and currently-deployed policies.

Looking at a time-series plot of allocations (\autoref{fig:mtadv:lc_vs_time}), we see that \Tagged performance improves over time, but fails to protect the IP pool against exploration for high $AR_{max}$. However, \Tagged still provides some protection even at these high allocation ratios, likely because it reduces the number of unique IPs with which tenants associate latent configuration. In contrast, \Segmented has an initial spike in exposed configurations, which then quickly stabilizes at a lower yield.

\paragraph{Effect of Tenant Count.} A realistic adversary may not have access to create an unlimited number of tenants in the public cloud, due to billing and other compliance measures taken by the provider. As such, it is helpful to understand how adversarial capability scales with number of tenants under various allocation policies. In \autoref{fig:mtadv:lc_vs_tc}, we see the marked effect of scaling tenant counts on effectiveness against \Tagged. An adversary begins to increase latent configuration yield above 20 tenants, with peak yields reached at 60 tenants. In contrast, \Segmented provides only a slight increase in yields even with no limit on tenants\footnote{A limit of $10^4$ in this scenario allows the adversary to never reuse a tenant, so the tenant count is effectively unlimited.}.

Our analysis of the multi-tenant adversary demonstrates the limitations of existing allocation policies, as an adversary using many tenants can still discover latent configuration. In contrast, IP scan segmentation's heuristics more effectively segment pool scanning based on the characteristics of allocations, rather than just tenant identifiers, and so are resistant to these attacks. Further, the \Segmented pool achieves improved performance even at very high pool contention, approaching the practical limit while maintaining existing minimum reuse durations.

\subsection{Tuning Segmentation}\label{sec:technique_eval:segmentation}
\begin{figure}[ht]
    \centering
    \includegraphics[width=\columnwidth]{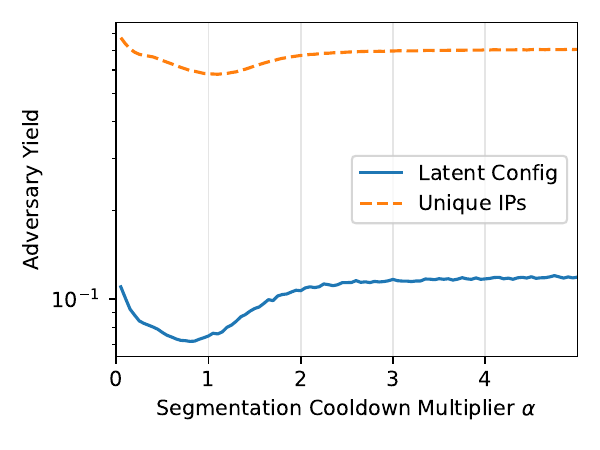}
    \caption{Effect of varying Segmentation parameter $\alpha$ on adversarial yield. ($AR_{max}=0.9$)}
    \label{fig:tuning_segmentation}
\end{figure}

Because the \Segmented policy is parameterized by some $\alpha$, it is important to tune the parameter for optimal performance. Here, we seek to understand how varying $\alpha$ affects adversarial yields under our simulation, and also how cloud providers might model policies on their own traces. To do this, we perform simulations of an unlimited-tenant adversary against a \Segmented pool. We vary $\alpha$ and study the yields of unique IPs and latent configuration. Simulation timelines are the same as in the multi-tenant evaluation (180 benign + 30 adversarial days), with 101 total experiments performed (58 total years of simulated allocation).

Our results (\autoref{fig:tuning_segmentation}) demonstrate the effect of varying $\alpha$. Varying $\alpha$ can make up to a \segmentedIPYieldVariation{} variation in unique IP yields, and up to a \segmentedLCYieldVariation{} variation in latent configuration yield. The relationship between $\alpha$ and adversarial objectives is convex, leading to a clear global optimum for configuration of a deployed system.

In addition to demonstrating an optimal value of $\alpha$ in our simulation setting, our results also suggest that modeling configurations of the \Segmented policy could be performed without making as strict of assumptions about latent configuration. Recall that our analysis assumes exponentially-distributed latent configuration durations. While a cloud provider could substitute real IP allocation traces, collecting data on concrete configurations is far more difficult. In our results, however, we show that latent configuration and IP yield are highly correlated, so a cloud provider could model IP address yields on concrete data and be confident in applicability to latent configuration yields, as well.

\begin{figure*}[t]
    \begin{subfigure}[t]{.32\textwidth}
        \centering
        \includegraphics[width=\textwidth]{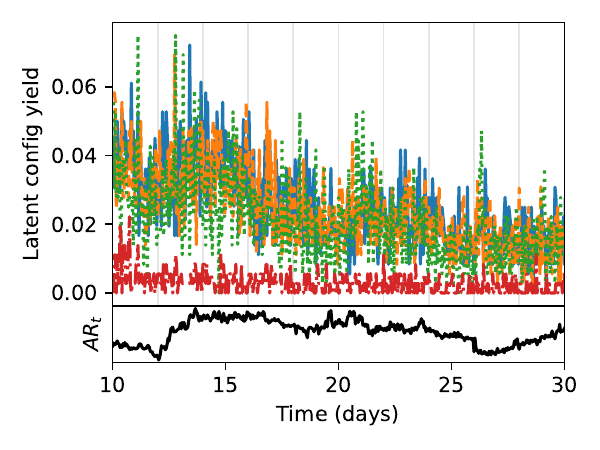}
        \caption{Yield of new latent configurations over time ($AR_{max}=0.95$).}
        \label{fig:borgadv:lc_vs_time}
    \end{subfigure}
    \hfill
    \begin{subfigure}[t]{.32\textwidth}
        \centering
    \includegraphics[width=\textwidth]{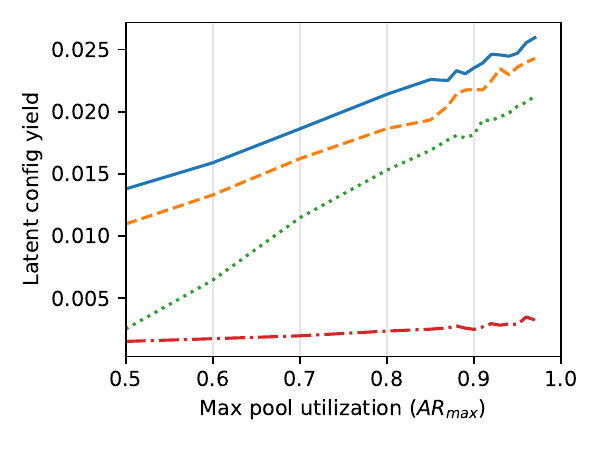}
    \caption{Effect of pool utilization on latent configurations. \Segmented protects real-world deployments even at high pool contention.}
    \label{fig:borgadv:lc_vs_ra}
    \end{subfigure}
    \hfill
     \begin{subfigure}[t]{.32\textwidth}
         \centering
        \includegraphics[width=\textwidth]{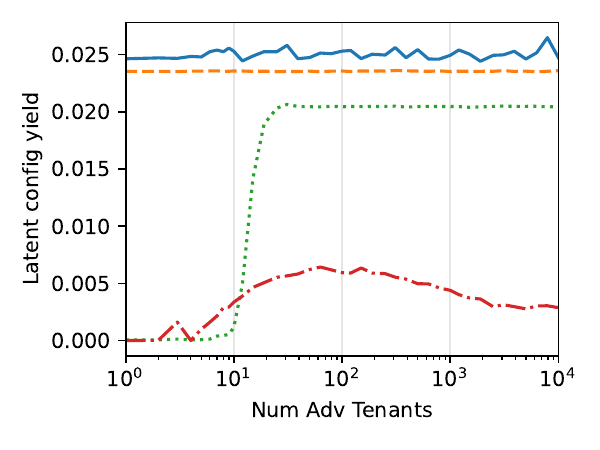}
        \caption{Effect of tenant count on latent configurations. \Segmented provides substantial protection against a multi-tenant adversary.}
         \label{fig:borgadv:lc_vs_tc}
    \end{subfigure}
    \centering
    \includegraphics[width=\columnwidth]{figs/legend.pdf}
    \caption{Evaluating allocation policies on real-world traces from \clusterdata.}
    \label{fig:borgadv}
\end{figure*}

\section{Model Realism}
\label{sec:realism}

We next demonstrate the realism of our model and policies by comparing our statistical models against data observed in real-world cloud settings (\autoref{sec:technique_eval:lc}), evaluating on real-world allocations (\autoref{sec:technique_eval:real}), and demonstrating the scalability of allocation policies to major cloud providers (\autoref{sec:discussion:policy_realism}). When evaluated in a realistic setting, IP Scan Segmentation performs even more favorably than under synthetic benchmarks. Further, our performance figures demonstrate that all evaluated policies can scale to the size and performance requirements of major providers.

\subsection{Validating Latent Configuration}\label{sec:technique_eval:lc}
\begin{figure}[t]
    \centering
    \includegraphics[width=\columnwidth]{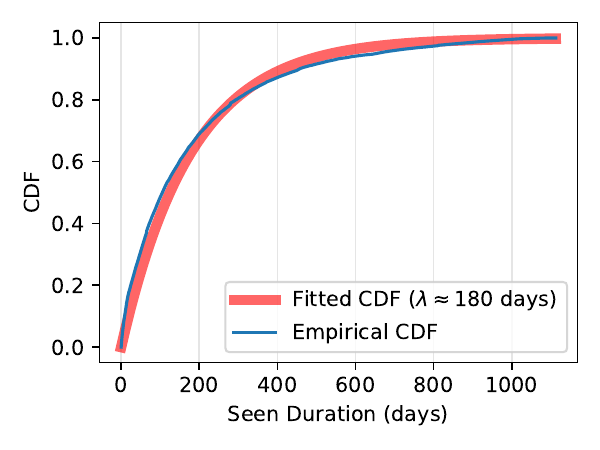}
    \caption{Distributions of latent configuration durations collected on real-world cloud traffic. Latent configuration durations fit to the hypothesized exponential distribution.}
    \label{fig:lc_duration}
\end{figure}

Finally, we evaluate the realism of our model of latent configuration by analyzing the distribution of latent configuration in deployed systems. To do this, we analyze real-world latent configurations as visible from a cloud-deployed Internet telescope~\cite{pauley_dscope_2023}. We use the DScope HTTP(S) request dataset spanning three years from March 2021-March 2024. Because many of the IP addresses studied are received many times, prevalence of latent configuration can be analyzed over time. For configurations seen multiple times in the study, we compute the maximum duration each configuration was seen, and use maximum likelihood estimation to fit a corresponding exponential distribution.

The resulting distribution (\autoref{fig:lc_duration}) is consistent with our hypothesized exponential distribution of latent configuration. While characterizing the distribution of latent configuration with respect to the underlying IP address allocation would require data from cloud providers, these empirical results support our statistical models and the effectiveness of our studied policies.

\subsection{Evaluating on Real Allocation}\label{sec:technique_eval:real}

We next evaluate whether our IP allocation model generalizes to real-world scenarios, testing allocation policies against server allocation traces from Google's \clusterdata dataset~\cite{google-cluster-data}. This dataset contains real-world server allocations and usage traces across 31 days in eight independent clusters in a hybrid cloud setting.

To extract corresponding IP address allocation traces from allocations in \clusterdata, we take all Job (groups of processes running as a single collection) allocations across all eight clusters, remove malformed jobs or those running beyond the scope of 31 days, and extract the User of these jobs as a tenant ID. Each Job is assumed to have a public IP address allocated, and latent configuration is modeled over these jobs as previously discussed. The resulting traces contain \SI{24}{M} allocations across \SI{21}{k} tenants, with $\max_{t}|\mathcal{I}_{A_t}| \approx \SI{119}{k}$.

Results (\autoref{fig:borgadv}) largely confirm the effectiveness of new IP allocation policies. Here, we see that \Segmented prevents discovery of latent configurations by an adversary with unlimited tenants, even at high pool utilization. Notably, \clusterdata{}'s composition of short-lived allocations for batch jobs represents a worst-case scenario, with many of these allocations seemingly indistinguishable from those used by an adversary. Yet, the \Segmented policy reduces the sharing of long-lived IP allocations with these short-lived tenants, preventing the adversary from discovering IPs with associated latent configuration.

One interesting phenomenon visible on these real-world allocation traces is the non-monotonic effect of tenant count on attack effectiveness. Here, we see that an adversary achieves increasing latent configuration yield with more tenants, then reduced effectiveness once tenants are no longer reused. This is a result of the default reputation of tenants: a new tenant has a $d_a/n_a$ of 0, which is then increased by allocating and releasing IPs. Reusing tenants with this (minimal) increase in reputation affords greater yield, especially when legitimate tenants have similar IP allocation behavior. While this worst-case scenario emphasizes a weakness of the \Segmented policy, it is unlikely that similar behavior would be seen in a public cloud, where job-based products such as AWS Lambda and Batch do not assign public IPs to short-lived instances.

\subsection{Performance \& Scalability}\label{sec:discussion:policy_realism}

\begin{table}[t]
    \caption{Performance scaling of IP Scan Segmentation with pool size. Speedup is the amount of simulated time (100 days) divided by time to simulate. IP Scan Segmentation scales to pools with millions of IPs and hundreds of millions of allocations.}
    \centering
    \begin{tabularx}{1.0\columnwidth}{r@{\extracolsep{\fill}}r@{\extracolsep{\fill}}r@{\extracolsep{\fill}}r@{\extracolsep{\fill}}r}
        \toprule
        \# IPs & Runtime & Speedup & Allocations & Allocs/s\\
        \midrule
        \SI{100}{} & \SI{500}{ms} & \SI{17}{M} & \SI{4.2}{k} & \SI{8.3}{k} \\ 
\SI{ 1}{k} & \SI{530}{ms} & \SI{16}{M} & \SI{26}{k} & \SI{48}{k} \\ 
\SI{10}{k} & \SI{ 2}{s} & \SI{4.3}{M} & \SI{220}{k} & \SI{110}{k} \\ 
\SI{100}{k} & \SI{14}{s} & \SI{630}{k} & \SI{2.2}{M} & \SI{160}{k} \\ 
\SI{ 1}{M} & \SI{187}{s} & \SI{46}{k} & \SI{22}{M} & \SI{120}{k} \\ 
\SI{10}{M} & \SI{2.3}{ks} & \SI{3.8}{k} & \SI{220}{M} & \SI{97}{k} \\ 

        \bottomrule
    \end{tabularx}
    \label{tab:perf_size}
\end{table}

\begin{table}[t]
    \caption{Largest major cloud compute regions.}
    \centering
    \begin{tabularx}{\columnwidth}{c@{\extracolsep{\fill}}c@{\extracolsep{\fill}}r@{\extracolsep{\fill}}r}
        \toprule
        Provider & Largest Region & \# Zones & \# IPs\\
        \midrule
        GCP~\cite{gcp_ip_ranges} & \texttt{us-central-1} & 4 & \SI{2.8}{M} \\
        Azure~\cite{azure_ip_ranges} & \texttt{eastus} & 3 & \SI{3.3}{M} \\
        AWS~\cite{aws_ip_ranges} & \texttt{us-east-1} & 5 & \SI{16}{M} \\
\bottomrule
    \end{tabularx}
    \label{tab:cloud_sizes}
\end{table}
Our work aims to provide practical security improvements through IP allocation policies, and it is therefore important that such policies are realizable. To this end, we evaluate the performance of IP Scan Segmentation on various IP address pool sizes. Notably, because \toolname simulates each concrete IP address allocation, the compute requirements required to simulate allocations are similar to that of a production allocation pool. We simulate non-adversarial scenarios on a commodity x64 server with 64vCPU and 192GB of RAM, though simulations use only one CPU thread. In each case, $|\mathcal{I}|/10$ tenants were used\footnote{The policies discussed store $O(1)$ data per tenant, so per-tenant compute overhead in \toolname is largely caused by simulating tenant agents, rather than the allocation policies themselves.} with a max concurrent allocation of $10$ per tenant. Simulations run for 100 (virtual) days. Results (\autoref{tab:perf_size}) show runtime and allocation rates with respect to pool size, demonstrating that \toolname scales with pool size to millions of allocations. The largest cloud compute regions (\autoref{tab:cloud_sizes}) can reallocate at most a few thousand addresses per second, well within the performance of \Segmented on a single CPU core.

Studied performance numbers align with the sizes of the largest cloud compute regions (see \autoref{tab:cloud_sizes}), demonstrating the proposed techniques scale to the size of major providers.

We further demonstrate the achievability of new policies by evaluating the real-world behavior of an existing provider and how those map to the information storage requirements of our proposed \Segmented policy. In the case of AWS, while allocation is random, AWS also already tags IP addresses with their previous tenant, and allows tenants to reuse released IPs if they have not been allocated to another tenant~\cite{aws_elastic_recover}. This currently-stored data is sufficient to perform the tenant tagging used by \Segmented and \Tagged policies. The remainder of the \Segmented policy requires associating an additional timestamp with each IP. Candidate IPs are then randomly sampled (as under current policies) and a best-fit IP is selected based on the heuristic. In this way, the \Segmented policy can be achieved using the existing data structures implemented by a major provider. 

\section{Limitations}
IP allocation policies are a heuristic mitigation, rather than a sound solution, for abuse of a cloud provider's IP address pool. Under our threat model, providers \textit{must} allocate some address to a tenant on request. Further, the provider will always face limited information, as a Sybil attack is not soundly distinguishable from benign new tenants.

\paragraph{Effects on Benign Tenants} While IP Scan Segmentation reduces the ability of adversaries to observe latent configuration and harm future customers \textit{in expectation}, it may also affect benign tenants, especially new customers. These new customers would be treated the same as new adversarial tenants, and may therefore receive disproportionately more low-quality IP addresses that were previously controlled by an adversary. Note that this applies only to prospective threats (i.e., a new benign tenant receiving an address that has been polluted by an adversary), as new tenants that then hold IPs for long periods will receive protection from retrospective threats.

To mitigate harms to new tenants, providers can add additional signals to the allocation process. The multi-tenant adversary requires access to many payment credentials that are likely of low quality (e.g., stolen credit cards), so countermeasures can privilege behaviors likely not associated with these. For instance, customers that purchase high-margin non-compute products (i.e., those not useful for adversarial IP allocation) or those with commercial contracts and vetted relationships, such as new accounts under existing billing arrangements. The effective price of leased addresses can also be considered (\autoref{sec:discussion:pricing}).

\paragraph{Provider Feedback and Implementation} Our proposed policies have not yet been deployed by major cloud providers. While our evaluation demonstrates that allocation policies can be effectively implemented from a technical perspective, other factors may prevent their adoption, such as associated reputational risks as providers take responsibility for client configurations. Cloud providers operate under a \textit{shared responsibility model}~\cite{alvarenga_what_2022}, wherein they take responsibility for infrastructure security and customers are responsible for their workloads. Defending against retrospective threats to IP allocation blurs this boundary, and potentially exposes providers to increased risk or scrutiny when protections fail. Similar actions have been taken in other shared domains, such as cloud storage security~\cite{claburn_aws_2022}, providing hope that evaluations of IP allocation effectiveness may lead to practical improvements in security.

Ultimately, we recommend that providers continue to embrace the shared responsibility model while protecting tenants when possible. In the case of IP address allocation, this would entail soft adoption of proposed allocation policies. Notably, all discussed policies are \textit{fully compatible} with the existing documented behavior of major providers. Non-documented adoption of new allocation policies would provide heuristic protection for customers while minimizing associated responsibility for the security of customer configurations.

\section{Discussion \& Related Work}\label{sec:discussion}

\subsection{Allocation Pricing Signals}\label{sec:discussion:pricing}

IP Scan Segmentation aims to increase the cost associated with IP scanning by tracking the amount spent per IP based on allocation time. In real cloud providers, this could motivate further extending the policy by incorporating other pricing signals from the cloud provider. For example, a tenant allocating powerful servers for short periods of batch processing is indistinguishable from scanning using just allocation traces, but the cloud provider could measure the total cost associated with these allocations and distinguish the activity as legitimate. The IP pool is a scarce resource, and so reducing the number of scanner-segmented IPs allocated to these resources will leave more available for scanners, improving policy effectiveness. We anticipate that cloud providers can extend the \toolname framework to incorporate these pricing signals and further improve practical security.

\subsection{Configuration Management}
While the security of IP address allocation has had limited study, the way that organizations \textit{configure} services has seen extensive related research. This is especially true in the cloud setting, which introduces new challenges in managing service configurations. Prior work has demonstrated that configuration complexity may increase substantially with the scale of the service~\cite{brown2004approach,tang2015holistic} and from the added tasks associated with making services cloud native (i.e., using advanced features such as auto-scaling~\cite{gannon2017cloud,dillon2010cloud}). Automated configuration management tools (such as Puppet~\cite{puppet}, Chef~\cite{progress_chef}, and Ansible~\cite{ansible}) have eased this complexity to some extent. Further, infrastructure-as-code (IaC)~\cite{guerriero2019adoption} tools (such as AWS CloudFormation~\cite{aws_cloudformation} or Terraform~\cite{terraform}) have made configuration management almost entirely non-interactive. However, while automation tools can eliminate most human-errors at runtime, a large proportion of configuration errors have been attributed to subtle bugs in the configuration files themselves (or ambiguities in the code generating them)~\cite{tang2015holistic} and other improper lifecycle management practices~\cite{brown2004approach} (e.g., failing to remove configurations pointing to released IPs~\cite{liu_all_2016, borgolte_cloud_2018, pauley_measuring_2022}).

\section{Conclusion}\label{sec:conclusion}
The way in which cloud IP addresses are allocated has a substantial impact on the security of hosted applications. Our work proposes new defenses for cloud IP allocation, and evaluates these defenses through a comprehensive model of tenant and adversarial behavior. Our proposed new policy, IP scan segmentation, successfully reduces an adversary's ability to scan the IP pool even if they can create new cloud tenants without limit. We anticipate that new IP allocation policies, such as IP scan segmentation, will prove useful to providers in protecting their customers and address pools. To that end, we release both our models and policies as open source for use by providers. We are also hopeful that modeling of IP allocation, such as that implemented in \toolname, will enable further improvements in the security of cloud provider offerings.

\section*{Acknowledgements}
The authors would like to thank the anonymous reviewers and our shepherd for their contribution towards improving this work. This material is based upon work supported by the National Science Foundation under Grant No. 2127200 and CNS-1900873, and by the National Science Foundation Graduate Research Fellowship Program under Grant No. DGE1255832. Any opinions, findings, and conclusions or recommendations expressed in this material are those of the authors and do not necessarily reflect the views of the National Science Foundation.

\appendices

\vfill

\begin{mdframed}
\section{Symbology}
    \centering
    \resizebox{\textwidth}{!}{
    \label{sec:symbols}
    \begin{tabular}{ll}
        \toprule
        \textbf{\textit{Symbol}} & \textbf{\textit{Meaning}}\\
        \midrule
        $a_i$                   & Fourier amplitude\\
        $AR_{max}$              & maximum pool allocation ratio\\
        $AR_t$                  & pool allocation ratio\\
        $\mathcal{B}$           & distribution of tenant behaviors\\
        $B_i$                   & behavior of tenant $i$\\
        $d$                     & duration\\
        $d_a$                   & duration of allocation\\
        $d_{reuse}$             & minimum time duration before an IP can be reused\\
        $d_{v}$                 & duration of vulnerability\\
        $\mathcal{I}$           & set of all IP addresses\\
        $\mathcal{I}_{A_{t}}$   & set of IP addresses currently allocated\\
        $p$                     & probability\\
        $p_c$                   & probability of latent configuration\\
        $S_{max}$               & maximum number of servers\\
        $S_{min}$               & minimum number of servers\\
        $t$                     & time \\
        $t_a$                   & allocation time\\
        $t_c$                   & time of configuration dissociation\\
        $t_{cd}$                & cooldown time\\
        $t_r$                   & release time\\
        $T$                     & tenant ID\\
        $\alpha$                & segmentation cooldown multiplier\\
        $\phi_i$                & Fourier phase\\
        $\theta$                & opaque state\\
        \bottomrule
    \end{tabular}}
    \vspace{10pt}
\end{mdframed}
\vspace{-0.59cm}
\hspace{2.5in}
\includegraphics[width=1cm]{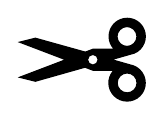}

\pagebreak[4]

{\AtNextBibliography{\small}\printbibliography}

@misc{aws_quotas,
  key     = {amazon},
  title   = {{AWS} service quotas - {AWS} {General} {Reference}},
  url     = {https://docs.aws.amazon.com/general/latest/gr/aws_service_limits.html},
  urldate = {2022-06-16}
}

@misc{aws_ip_ranges,
  title   = {{AWS} {IP} address ranges - {AWS} {General} {Reference}},
  url     = {https://docs.aws.amazon.com/general/latest/gr/aws-ip-ranges.html},
  urldate = {2021-08-16},
  key     = {aws}
}

@misc{gcp_ip_ranges,
  title = {{GCP} {IP} address ranges},
  url   = {https://www.gstatic.com/ipranges/cloud.json},
  key   = {gcp}
}

@misc{azure_ip_ranges,
  title = {{Azure} {IP} address ranges},
  url   = {https://www.microsoft.com/en-us/download/details.aspx?id=56519},
  key   = {azure}
}

@techreport{holdrege_ip_1999,
  type        = {Request for {Comments}},
  title       = {{IP} {Network} {Address} {Translator} ({NAT}) {Terminology} and {Considerations}},
  url         = {https://datatracker.ietf.org/doc/rfc2663},
  number      = {RFC 2663},
  urldate     = {2022-06-10},
  institution = {Internet Engineering Task Force},
  author      = {Holdrege, Matt and Srisuresh, Pyda},
  month       = aug,
  year        = {1999},
  doi         = {10.17487/RFC2663},
  note        = {Num Pages: 30}
}

@techreport{droms_dynamic_1997,
  type        = {Request for {Comments}},
  title       = {Dynamic {Host} {Configuration} {Protocol}},
  url         = {https://datatracker.ietf.org/doc/rfc2131},
  number      = {RFC 2131},
  urldate     = {2022-06-10},
  institution = {Internet Engineering Task Force},
  author      = {Droms, Ralph},
  month       = mar,
  year        = {1997},
  doi         = {10.17487/RFC2131},
  note        = {Num Pages: 45}
}

@inproceedings{liu_all_2016,
  address    = {Vienna Austria},
  title      = {All {Your} {DNS} {Records} {Point} to {Us}: {Understanding} the {Security} {Threats} of {Dangling} {DNS} {Records}},
  isbn       = {978-1-4503-4139-4},
  shorttitle = {All {Your} {DNS} {Records} {Point} to {Us}},
  url        = {https://dl.acm.org/doi/10.1145/2976749.2978387},
  doi        = {10.1145/2976749.2978387},
  language   = {en},
  urldate    = {2020-09-14},
  booktitle  = {Proceedings of the 2016 {ACM} {SIGSAC} {Conference} on {Computer} and {Communications} {Security}},
  publisher  = {ACM},
  author     = {Liu, Daiping and Hao, Shuai and Wang, Haining},
  month      = oct,
  year       = {2016},
  pages      = {1414--1425}
}

@inproceedings{borgolte_cloud_2018,
  address    = {San Diego, CA},
  title      = {Cloud {Strife}: {Mitigating} the {Security} {Risks} of {Domain}-{Validated} {Certificates}},
  isbn       = {978-1-891562-49-5},
  shorttitle = {Cloud {Strife}},
  url        = {https://www.ndss-symposium.org/wp-content/uploads/2018/02/ndss2018_06A-4_Borgolte_paper.pdf},
  doi        = {10.14722/ndss.2018.23327},
  language   = {en},
  urldate    = {2021-08-04},
  booktitle  = {Proceedings 2018 {Network} and {Distributed} {System} {Security} {Symposium}},
  publisher  = {Internet Society},
  author     = {Borgolte, Kevin and Fiebig, Tobias and Hao, Shuang and Kruegel, Christopher and Vigna, Giovanni},
  year       = {2018}
}

@inproceedings{pauley_measuring_2022,
  title     = {Measuring and {Mitigating} the {Risk} of {IP} {Reuse} on {Public} {Clouds}},
  isbn      = {978-1-66541-316-9},
  url       = {https://www.computer.org/csdl/proceedings-article/sp/2022/131600b523/1CIO7rpcgSs},
  doi       = {10.1109/SP46214.2022.00094},
  language  = {English},
  urldate   = {2022-05-28},
  booktitle = {2022 {IEEE} {Symposium} on {Security} and {Privacy} ({SP})},
  publisher = {IEEE Computer Society},
  author    = {Pauley, Eric and Sheatsley, Ryan and Hoak, Blaine and Burke, Quinn and Beugin, Yohan and McDaniel, Patrick},
  month     = apr,
  year      = {2022},
  note      = {ISSN: 2375-1207},
  pages     = {1523--1523}
}

@techreport{moskowitz_address_1996,
  type        = {Request for {Comments}},
  title       = {Address {Allocation} for {Private} {Internets}},
  url         = {https://datatracker.ietf.org/doc/rfc1918},
  number      = {RFC 1918},
  urldate     = {2022-06-07},
  institution = {Internet Engineering Task Force},
  author      = {Moskowitz, Robert and Karrenberg, Daniel and Rekhter, Yakov and Lear, Eliot and Groot, Geert Jan de},
  month       = feb,
  year        = {1996},
  doi         = {10.17487/RFC1918},
  note        = {Num Pages: 9}
}

@article{noauthor_ipv6_nodate,
  author   = {Strowes, Stephen},
  title    = {{IPv6} {Adoption} in 2021},
  url      = {https://labs.ripe.net/author/stephen_strowes/ipv6-adoption-in-2021/},
  language = {en-US},
  urldate  = {2021-12-01},
  year     = {2021},
  journal  = {RIPE Labs}
}

@misc{alvarenga_what_2022,
  title      = {What is the {Shared} {Responsibility} {Model}? - {CrowdStrike}},
  shorttitle = {What is the {Shared} {Responsibility} {Model}?},
  url        = {https://www.crowdstrike.com/cybersecurity-101/cloud-security/shared-responsibility-model/},
  language   = {en},
  urldate    = {2024-07-14},
  journal    = {crowdstrike.com},
  author     = {Alvarenga, Gui},
  month      = nov,
  year       = {2022}
}

@misc{aws_autoscale_termination,
  key     = {Amazon},
  title   = {Work with {Amazon} {EC2} {Auto} {Scaling} termination policies - {Amazon} {EC2} {Auto} {Scaling}},
  url     = {https://docs.aws.amazon.com/autoscaling/ec2/userguide/ec2-auto-scaling-termination-policies.html},
  urldate = {2022-06-09}
}

@misc{sanfilippo_random_nodate,
  title   = {Random notes on improving the {Redis} {LRU} algorithm},
  url     = {http://antirez.com/news/109},
  urldate = {2022-06-14},
  author  = {Sanfilippo, Salvatore}
}

@inproceedings{tang2015holistic,
  title     = {Holistic configuration management at facebook},
  author    = {Tang, Chunqiang and Kooburat, Thawan and Venkatachalam, Pradeep and Chander, Akshay and Wen, Zhe and Narayanan, Aravind and Dowell, Patrick and Karl, Robert},
  booktitle = {Proceedings of the 25th Symposium on Operating Systems Principles},
  pages     = {328--343},
  year      = {2015}
}

@inproceedings{guerriero2019adoption,
  title        = {Adoption, support, and challenges of infrastructure-as-code: Insights from industry},
  author       = {Guerriero, Michele and Garriga, Martin and Tamburri, Damian A and Palomba, Fabio},
  booktitle    = {2019 IEEE International Conference on Software Maintenance and Evolution (ICSME)},
  pages        = {580--589},
  year         = {2019},
  organization = {IEEE}
}

@inproceedings{dillon2010cloud,
  title        = {Cloud computing: issues and challenges},
  author       = {Dillon, Tharam and Wu, Chen and Chang, Elizabeth},
  booktitle    = {2010 24th IEEE international conference on advanced information networking and applications},
  pages        = {27--33},
  year         = {2010},
  organization = {Ieee}
}

@inproceedings{brown2004approach,
  title     = {An approach to benchmarking configuration complexity},
  author    = {Brown, Aaron B and Hellerstein, Joseph L},
  booktitle = {Proceedings of the 11th workshop on ACM SIGOPS European workshop},
  pages     = {18--es},
  year      = {2004}
}

@inproceedings{herbst2013elasticity,
  title     = {Elasticity in cloud computing: What it is, and what it is not},
  author    = {Herbst, Nikolas Roman and Kounev, Samuel and Reussner, Ralf},
  booktitle = {10th International Conference on Autonomic Computing (ICAC 13)},
  pages     = {23--27},
  year      = {2013}
}

@inproceedings{chaisiri2011cost,
  title        = {Cost minimization for provisioning virtual servers in amazon elastic compute cloud},
  author       = {Chaisiri, Sivadon and Kaewpuang, Rakpong and Lee, Bu-Sung and Niyato, Dusit},
  booktitle    = {2011 IEEE 19th Annual International Symposium on Modelling, Analysis, and Simulation of Computer and Telecommunication Systems},
  pages        = {85--95},
  year         = {2011},
  organization = {IEEE}
}

@article{hwang2013cost,
  title     = {Cost optimization of elasticity cloud resource subscription policy},
  author    = {Hwang, Ren-Hung and Lee, Chung-Nan and Chen, Yi-Ru and Zhang-Jian, Da-Jing},
  journal   = {IEEE Transactions on Services Computing},
  volume    = {7},
  number    = {4},
  pages     = {561--574},
  year      = {2013},
  publisher = {IEEE}
}

@inproceedings{he2013next,
  title     = {Next stop, the cloud: Understanding modern web service deployment in ec2 and azure},
  author    = {He, Keqiang and Fisher, Alexis and Wang, Liang and Gember, Aaron and Akella, Aditya and Ristenpart, Thomas},
  booktitle = {Proceedings of the 2013 conference on Internet measurement conference},
  pages     = {177--190},
  year      = {2013}
}

@article{wolke2015more,
  title     = {More than bin packing: Dynamic resource allocation strategies in cloud data centers},
  author    = {Wolke, Andreas and Tsend-Ayush, Boldbaatar and Pfeiffer, Carl and Bichler, Martin},
  journal   = {Information Systems},
  volume    = {52},
  pages     = {83--95},
  year      = {2015},
  publisher = {Elsevier}
}

@article{bhavani2014resource,
  title   = {Resource provisioning techniques in cloud computing environment: a survey},
  author  = {Bhavani, BH and Guruprasad, HS},
  journal = {International Journal of Research in Computer and Communication Technology},
  volume  = {3},
  number  = {3},
  pages   = {395--401},
  year    = {2014}
}

@inproceedings{calheiros2011virtual,
  title        = {Virtual machine provisioning based on analytical performance and QoS in cloud computing environments},
  author       = {Calheiros, Rodrigo N and Ranjan, Rajiv and Buyya, Rajkumar},
  booktitle    = {2011 International Conference on Parallel Processing},
  pages        = {295--304},
  year         = {2011},
  organization = {IEEE}
}

@article{qu2018auto,
  title     = {Auto-scaling web applications in clouds: A taxonomy and survey},
  author    = {Qu, Chenhao and Calheiros, Rodrigo N and Buyya, Rajkumar},
  journal   = {ACM Computing Surveys (CSUR)},
  volume    = {51},
  number    = {4},
  pages     = {1--33},
  year      = {2018},
  publisher = {ACM New York, NY, USA}
}

@article{gannon2017cloud,
  title     = {Cloud-native applications},
  author    = {Gannon, Dennis and Barga, Roger and Sundaresan, Neel},
  journal   = {IEEE Cloud Computing},
  volume    = {4},
  number    = {5},
  pages     = {16--21},
  year      = {2017},
  publisher = {IEEE}
}

@misc{yuan_scryer_2017,
  title      = {Scryer: {Netflix}’s {Predictive} {Auto} {Scaling} {Engine} — {Part} 2},
  shorttitle = {Scryer},
  url        = {https://netflixtechblog.com/scryer-netflixs-predictive-auto-scaling-engine-part-2-bb9c4f9b9385},
  language   = {en},
  urldate    = {2022-06-15},
  journal    = {Netflix Technology Blog},
  author     = {Yuan, Danny and Joshi, Neeraj and Jacobson, Daniel and Oberai, Puneet},
  month      = apr,
  year       = {2017}
}

@misc{aws_website,
  key      = {Amazon},
  title    = {Cloud {Services} - {Amazon} {Web} {Services} ({AWS})},
  url      = {https://aws.amazon.com/},
  language = {en-US},
  urldate  = {2021-08-19},
  journal  = {Amazon Web Services, Inc.}
}

@misc{google_cloud_website,
  key     = {Google},
  title   = {Cloud {Computing} {Services}  {\textbar}  {Google} {Cloud}},
  url     = {https://cloud.google.com/},
  urldate = {2021-08-19}
}

@misc{azure_website,
  key      = {Microsoft},
  title    = {Cloud {Computing} {Services} {\textbar} {Microsoft} {Azure}},
  url      = {https://azure.microsoft.com/en-us/},
  language = {en},
  urldate  = {2021-08-19}
}

@misc{progress_chef,
  key   = {Progress},
  title = {Progress {Chef}},
  url   = {https://www.chef.io/}
}

@misc{ansible,
  key   = {Ansible},
  title = {Ansible},
  url   = {https://www.ansible.com/}
}

@misc{puppet,
  key   = {Puppet},
  title = {Puppet},
  url   = {https://puppet.com/}
}

@misc{aws_cloudformation,
  key   = {Amazon},
  title = {AWS {CloudFormation}},
  url   = {https://aws.amazon.com/cloudformation/}
}

@misc{terraform,
  key   = {HashiCorp},
  title = {Terraform by {HashiCorp}},
  url   = {https://www.terraform.io/}
}

@inproceedings{google-cluster-data,
  title     = {Large-scale cluster management at {Google} with {Borg}},
  author    = {Abhishek Verma and Luis Pedrosa and Madhukar R. Korupolu and David Oppenheimer and Eric Tune and John Wilkes},
  year      = {2015},
  booktitle = {Proceedings of the European Conference on Computer Systems (EuroSys)},
  address   = {Bordeaux, France}
}

@misc{aws_elastic_recover,
  title   = {Elastic {IP} addresses - {Amazon} {Elastic} {Compute} {Cloud}},
  url     = {https://docs.aws.amazon.com/AWSEC2/latest/UserGuide/elastic-ip-addresses-eip.html#using-eip-recovering},
  urldate = {2023-02-13}
}

@article{almohri_predictability_2020,
  title    = {Predictability of {IP} {Address} {Allocations} for {Cloud} {Computing} {Platforms}},
  volume   = {15},
  issn     = {1556-6021},
  doi      = {10.1109/TIFS.2019.2924555},
  journal  = {IEEE Transactions on Information Forensics and Security},
  author   = {Almohri, Hussain M. J. and Watson, Layne T. and Evans, David},
  year     = {2020},
  note     = {Conference Name: IEEE Transactions on Information Forensics and Security},
  pages    = {500--511}
}

@article{shannon1949communication,
  title     = {Communication in the presence of noise},
  author    = {Shannon, Claude E},
  journal   = {Proceedings of the IRE},
  volume    = {37},
  number    = {1},
  pages     = {10--21},
  year      = {1949},
  publisher = {IEEE}
}

@misc{spamhaus,
  title    = {Who is {Spamhaus} - the leader in {IP} \& domain reputation data},
  url      = {https://www.spamhaus.org/who-is-spamhaus/},
  language = {en},
  journal  = {The Spamhaus Project}
}

@inproceedings{ioannidis_implementing_2000,
  address   = {Athens Greece},
  title     = {Implementing a distributed firewall},
  isbn      = {978-1-58113-203-8},
  url       = {https://dl.acm.org/doi/10.1145/352600.353052},
  doi       = {10.1145/352600.353052},
  language  = {en},
  booktitle = {Proceedings of the 7th {ACM} conference on {Computer} and {Communications} {Security}},
  publisher = {ACM},
  author    = {Ioannidis, Sotiris and Keromytis, Angelos D. and Bellovin, Steve M. and Smith, Jonathan M.},
  month     = nov,
  year      = {2000},
  pages     = {190--199}
}

@inproceedings{pauley_dscope_2023,
  address   = {Anaheim, CA},
  title     = {{DScope}: {A} {Cloud}-{Native} {Internet} {Telescope}},
  booktitle = {Proceedings of the 32nd {USENIX} {Security} {Symposium} ({USENIX} {Security} 2023, to appear)},
  publisher = {USENIX Association},
  author    = {Pauley, Eric and Barford, Paul and McDaniel, Patrick},
  month     = aug,
  year      = {2023}
}

@misc{claburn_aws_2022,
  title    = {{AWS} simplifies {Simple} {Storage} {Service} to prevent data leaks},
  url      = {https://www.theregister.com/2022/12/14/aws_simple_storage_service_simplified/},
  language = {en},
  journal  = {The Register},
  author   = {Claburn, Thomas},
  month    = dec,
  year     = {2022}
}

@inproceedings{sinha_shades_2008,
  address    = {Alexandria, VA, USA},
  title      = {Shades of grey: {On} the effectiveness of reputation-based blacklists},
  isbn       = {978-1-4244-3288-2},
  shorttitle = {Shades of grey},
  url        = {http://ieeexplore.ieee.org/document/4690858/},
  doi        = {10.1109/MALWARE.2008.4690858},
  language   = {en},
  booktitle  = {2008 3rd {International} {Conference} on {Malicious} and {Unwanted} {Software} ({MALWARE})},
  publisher  = {IEEE},
  author     = {Sinha, Sushant and Bailey, Michael and Jahanian, Farnam},
  month      = oct,
  year       = {2008},
  pages      = {57--64}
}

@inproceedings{zhang_highly_2008,
  title     = {Highly predictive blacklisting.},
  url       = {https://www.usenix.org/legacy/events/sec08/tech/full_papers/zhang/zhang.pdf},
  booktitle = {{USENIX} security symposium},
  author    = {Zhang, Jian and Porras, Phillip A. and Ullrich, Johannes},
  year      = {2008},
  pages     = {107--122}
}

@inproceedings{antonakakis_building_2010,
  title     = {Building a dynamic reputation system for {DNS}},
  url       = {https://www.usenix.org/event/sec10/tech/full_papers/Antonakakis.pdf},
  booktitle = {19th {USENIX} {Security} {Symposium} ({USENIX} {Security} 10)},
  author    = {Antonakakis, Manos and Perdisci, Roberto and Dagon, David and Lee, Wenke and Feamster, Nick},
  year      = {2010}
}

@misc{howard_networking_2020,
  title      = {Networking {Education}: {IP} {Blocklist} \& {Removal}},
  shorttitle = {Networking {Education}},
  url        = {https://ipv4.global/blog/ip-blocklist-blacklist/},
  language   = {en},
  journal    = {IPv4 Global},
  author     = {Howard, Lee},
  month      = jun,
  year       = {2020}
}

@inproceedings{esquivel_effectiveness_2010,
  title     = {On the effectiveness of {IP} reputation for spam filtering},
  url       = {https://ieeexplore.ieee.org/abstract/document/5431981},
  doi       = {10.1109/COMSNETS.2010.5431981},
  booktitle = {2010 {Second} {International} {Conference} on {COMmunication} {Systems} and {NETworks} ({COMSNETS} 2010)},
  author    = {Esquivel, Holly and Akella, Aditya and Mori, Tatsuya},
  month     = jan,
  year      = {2010},
  note      = {ISSN: 2155-2509},
  pages     = {1--10}
}

@misc{gardner_how_2023,
  title    = {How adversaries infiltrate {AWS} cloud accounts},
  url      = {https://redcanary.com/blog/aws-sts/},
  language = {en},
  journal  = {Red Canary},
  author   = {Gardner, Thomas and Betsworth, Cody},
  year     = {2023}
}

@misc{goodin_developers_2023,
  title    = {Developers can’t seem to stop exposing credentials in publicly accessible code},
  url      = {https://arstechnica.com/security/2023/11/developers-cant-seem-to-stop-exposing-credentials-in-publicly-accessible-code/},
  language = {en-us},
  journal  = {Ars Technica},
  author   = {Goodin, Dan},
  month    = nov,
  year     = {2023}
}

@misc{page_microsoft_2024,
  title    = {Microsoft employees exposed internal passwords in security lapse},
  url      = {https://techcrunch.com/2024/04/09/microsoft-employees-exposed-internal-passwords-security-lapse/},
  language = {en-US},
  journal  = {TechCrunch},
  author   = {Page, Zack Whittaker {and} Carly},
  month    = apr,
  year     = {2024}
}

@software{eric_pauley_2024_13698654,
  author    = {Eric Pauley},
  title     = {MadSP-McDaniel/eipsim: NDSS Publication},
  month     = sep,
  year      = 2024,
  publisher = {Zenodo},
  version   = {ndss-artifact},
  doi       = {10.5281/zenodo.13698654},
  url       = {https://doi.org/10.5281/zenodo.13698654}
}

\section{Artifact Appendix}
\noindent
This appendix describes artifacts to reproduce paper results.\footnote{Section V.A (Validating Latent Configuration) is a validation step performed using third-party data and is excluded from the paper artifacts.}

\subsection{Description \& Requirements}
\subsubsection{How to access}
Source code is publicly available~\cite{eric_pauley_2024_13698654}.

\subsubsection{Hardware dependencies}
Tested on a 48-vCPU x86 machine. More cores will speed up analysis results (4GB of RAM per vCPU recommended).

\subsubsection{Software dependencies} 
Linux, Go 1.18 or higher and Python 3. A Dockerfile is provided to install dependencies.

\subsubsection{Benchmarks} 
Server allocation traces from the \texttt{clusterdata-2019} dataset are required for real-world evaluations. These are included in the repository.

\subsection{Artifact Installation \& Configuration}

This section should include all the high-level installation and configuration steps required to prepare the environment to be used for the evaluation of your artifact.

\subsection{Experiment Workflow}\label{workflow}

All dependencies can be installed using Docker:

\noindent
\texttt{docker build -t eipsim .\\
docker run -it eipsim bash
}

Experiments are written as Go tests, with result visualization performed in Python. All tests and benchmarks can be performed at once using the following command (or by running specific tests defined under major claims):

\noindent
\texttt{go test -v ./eval/... -run . -bench . --timeout 10000m}

Results are stored as JSON files that can be manually inspected, or all paper results can be plotted:

\noindent
\texttt{cd eval \&\& python3 figs.py}

The figure code will automatically plot figures for which data is available, and skip figures with incomplete data.

\subsection{Major Claims}

\begin{itemize}
    \item (C1): As shown in Figure 4, the \textsc{Tagged} and \textsc{Segmented} policies reduce the prevalence of latent configuration collected by benign tenants.
    \item (C2): When an adversary uses many cloud tenants to exploit IP allocation (Figure 5) the \textsc{Segmented} policy reduces yield of latent configuration by adversaries across pool utilization and tenant count.
    \item (C3): Tuning the \textsc{Segmented} policy (Figure 6) further reduces adversarial yield of latent configuration.
    \item (C4): The \textsc{Segmented} policy can be implemented with performance characteristics (Table 1) acceptable to major providers.
    \item (C5): On real-world traces (Figure 8), the \textsc{Segmented} policy reduces adversarial yield of latent configuration compared to existing approaches.
\end{itemize}

\subsection{Evaluation}
Each individual experiment is a Go test, and can be run as follows:

\noindent
\texttt{go test -v ./eval/... -run [TestName] --timeout 10000m}

For convenience, it is recommended to run all tests at once as described in \autoref{workflow}.

All tests durations are defined in vCPU-hours. The tests will automatically parallelize across available vCPUs. After performing tests, results can be plotted using:

\texttt{cd eval \&\& python3 figs.py}

\subsubsection{Experiment (E1)}
[\texttt{TestBenign}] [5 human-minutes + 1 vCPU-hour]: policy performance under the benign scenario (C1)

\textit{[Execution]}
\texttt{go test -v ./eval/... -run TestBenign -timeout 10000m}

\textit{[Results]}
The resulting figure (\texttt{syn-benign-LC\_vs\_RA}) shows that the \textsc{Tagged} and \textsc{Segmented} policies reduce prevalence of latent configuration across pool utilization.

\subsubsection{Experiment (E2)}
[\texttt{TestAdversaryAgainstPoolPolicies}] [5 human-minutes + 200 vCPU-hour]: policy performance under multi-tenant adversary (C2)

\textit{[Execution]}
\texttt{go test -v ./eval/... -run TestAdversaryAgainstPoolPolicies -timeout 10000m}

\textit{[Results]}
Results over pool utilization (\texttt{syn-mtadv-newLatentConfs\_vs\_RA}) shows that the \textsc{Segmented} policy prevents adversarial exploitation across all allocation ratios. Results over adversarial tenant counts (\texttt{syn-newLatentConfs\_vs\_TC}) demonstrate this across adversarial tenant counts.

\subsubsection{Experiment (E3)}
[\texttt{TestSegmentedPoolSize}] [5 human-minutes + 50 vCPU-hour]: tests parameters for the \textsc{Segmented} policy (C3)

\textit{[Execution]}
\texttt{go test -v ./eval/... -run TestSegmentedPoolSize -timeout 10000m}

\textit{[Results]}
Results (\texttt{segmented\_multipliers}) show that tuning segmentation reduces adversarial yield of unique IPs and latent configuration.

\subsubsection{Experiment (E4)}
[\texttt{BenchmarkSimPerfSizeTest}] [5 human-minutes + 1 vCPU-hour]: tests performance of the \textsc{Segmented} policy (C4)

\textit{[Execution]}
\texttt{go test -v ./eval/... -bench BenchmarkSimPerfSizeTest -timeout 10000m}

\textit{[Results]}
Benchmark results show similar performance to that in the paper, with throughput acceptable for deployment on major providers.

\subsubsection{Experiment (E5)}
[\texttt{TestBorgAdversaries}] [5 human-minutes + 50 vCPU-hour]: tests policy performance on a real-world workload (C5)

\textit{[Execution]}
\texttt{go test -v ./eval/... -bench TestBorgAdversaries -timeout 10000m}

\textit{[Results]}
Results (\texttt{borg-mtadv-newLatentConfs\_vs\_RA}) show improved performance from the \textsc{Segmented} policy on real-world workloads.

\end{document}